\documentclass[twocolumn]{aastex63}
\usepackage{amsmath}

\newcommand{\harm}{\textsc{Harm3d} }

\def\gmunu{g_{\mu \nu}} \def\etamunu{\eta_{\mu \nu}} \def\lmu{l_{\mu}} \def\lnu{l_{\nu}}
  \def\xalpha{x^{\alpha}} \def\xbeta{x^{\beta}}
 \def\Lambdaalphabeta{{\Lambda^{\alpha}}_{\beta}}
  \def\xbaralpha{\bar{x}^{\alpha}} \def\xbarbeta{\bar{x}^{\beta}}

\def\pmag{p_{\mathrm{m}}}
\def\Lcool{\mathcal{L}_{\mathrm{cool}}}
\providecommand{\e}[1]{\ensuremath{\times 10^{#1}}}
\providecommand{\eqref}[1]{(\ref{#1})}
\def\runmspins{\texttt{b20-spins}}
\def\runvz{\texttt{b20\_v0}}
\def\runvo{\texttt{b20\_v1}}
\def\runvt{\texttt{b20\_v2}}
\def\runpspins{\texttt{b20+spins}}

\submitjournal{ApJ}

\shorttitle{Circumbinary Disk Accretion into Spinning Black Hole Binaries}
\shortauthors{Lopez Armengol, et al. (2021)}

\begin{document}

\title{Circumbinary Disk Accretion into Spinning Black Hole Binaries}

\correspondingauthor{Federico~G.~Lopez~Armengol}
\email{fglsma@rit.edu}
\author[0000-0002-4882-5672]{Federico~G.~Lopez~Armengol}
\affiliation{
Center for Computational Relativity and Gravitation, Rochester Institute of Technology \\
Rochester, NY 14623, USA}
\affiliation{Instituto Argentino de Radioastronom\'ia (CONICET; CICPBA) \\
C.C. No. 5, 1894 Villa Elisa, Argentina}
\author[0000-0002-5427-1207]{Luciano~Combi}
\affiliation{Instituto Argentino de Radioastronom\'ia (CONICET; CICPBA) \\
C.C. No. 5, 1894 Villa Elisa, Argentina}
\affiliation{
Center for Computational Relativity and Gravitation, Rochester Institute of Technology \\
Rochester, NY 14623, USA}
\author[0000-0002-8659-6591]{Manuela~Campanelli}
\affiliation{
Center for Computational Relativity and Gravitation, Rochester Institute of Technology \\
Rochester, NY 14623, USA}
\author[0000-0003-3547-8306]{Scott~C.~Noble}
\affiliation{Gravitational Astrophysics Laboratory, NASA Goddard Space Flight Center \\
Greenbelt, MD 20771, USA}
\author[0000-0002-2995-7717]{Julian~H.~Krolik}
\affiliation{Physics and Astronomy Department, Johns Hopkins University \\
Baltimore, MD 21218, USA}
\author[0000-0002-7447-1142]{Dennis~B.~Bowen}
\affiliation{X Computational Physics, Los Alamos National Laboratory \\
Los Alamos, New Mexico 87545}
\author[0000-0001-9562-9677]{Mark~J.~Avara}
\affiliation{Institute of Astronomy, University of Cambridge \\
Madingley Road, CB3 0HA Cambridge, UK}
\affiliation{
Center for Computational Relativity and Gravitation, Rochester Institute of Technology \\
Rochester, NY 14623, USA}
\author[0000-0001-5869-8542]{Vassilios~Mewes}
\affiliation{National Center for Computational Sciences, Oak Ridge National Laboratory \\
P.O. Box 2008, Oak Ridge, TN 37831-6164, USA}
\affiliation{Physics Division, Oak Ridge National Laboratory \\
P.O. Box 2008, Oak Ridge, TN 37831-6354, USA}
\affiliation{
Center for Computational Relativity and Gravitation, Rochester Institute of Technology \\
Rochester, NY 14623, USA}
\author[0000-0001-7665-0796]{Hiroyuki~Nakano}
\affiliation{Faculty of Law, Ryukoku University \\
Kyoto 612-8577, Japan}



\begin{abstract}

Supermassive black hole binaries are likely to accrete interstellar gas through a circumbinary disk.
Shortly before merger, the inner portions of this circumbinary disk are subject to general relativistic effects.
To study this regime, we approximate the spacetime metric of close orbiting black holes by superimposing two boosted Kerr-Schild terms.
After demonstrating the quality of  this approximation, we carry out very long-term general relativistic magnetohydrodynamic simulations of the circumbinary disk.
We consider black holes with spin dimensionless parameters of magnitude 0.9, in one simulation parallel to the orbital angular momentum of the binary, but in another anti-parallel.
These are contrasted with spinless simulations.
We find that, for a fixed  surface mass density in the inner circumbinary disk, aligned spins of this magnitude approximately reduce the mass accretion rate by 14\% and counter-aligned spins increase it by 45\%, leaving many other disk properties unchanged.

\end{abstract}



\section{Introduction} \label{sec:introduction}
The existence of
supermassive binary black holes (SMBBH) is a natural prediction of the
current hierarchical models of galaxy formation \citep{Merritt2005}.
After two galaxies merge, there are reasons to think the orbit of the
newly formed binary will shrink to a sub-parsec scale by dynamical
friction and interaction with surrounding gas \citep{Begelman1980,
Mayer2007, Escala2004, Escala2005, Merrit2004, Merritt2006, Dotti2007,
Dotti2009b, Shi2012,SesanaKhan2015,Mirza2017,Khan2019, Tiede+2020}.
From then on, the emission of gravitational
waves becomes an efficient mechanism for energy and angular momentum
extraction until coalescence \citep{PhysRevLett.95.121101,
PhysRevLett.96.111101, PhysRevLett.96.111102}. This emission of  gravitational waves makes SMBBHs the primary targets in the mHZ frequency window by
the future Laser Interferometer Space Antenna
\citep[LISA,][]{LISA2017} and by pulsar timing techniques in
the nHz range \citep{Nanograv2020, EPTA2016, Reardon2015}.

Unlike stellar-mass binary black hole systems, the environment of SMBBHs
might be rich in gas \citep{Barnes1992, Barnes1996, Mihos1996,
Mayer2007, Dotti2012, Mayer2013, Derdzinski2019}, allowing the system to
emit electromagnetic (EM) radiation. Many signatures have been proposed
as ways to hunt for SMBBHs in the EM spectrum: periodic light curves in
active galactic nuclei \citep[AGN;][]{Valtonen2006, Graham2015a,
Graham2015b, Liu2019,Saade2020}, interrupted jet activity
\citep{Shoenmakers2000, Liu2003}, traces of jet precession or
``spin-flips'' in X-shaped radio galaxies \citep{Merritt2002}, dual
compact radio cores \citep{Rodriguez2006}, shifts in the profiles of
broad emission lines \citep{Dotti2009, Bogdanovic2009}, X-ray emission
from streams striking the accretion disks around the individual black
holes or  a ``notch'' in the optical/IR spectrum \citep{Roedig2014,
Krolik2019}. However, it is not at all clear whether any of these
can be truly expected.
Numerical simulations are key guides to this
search because they may unveil unique dynamics and radiative
properties.

Matter flows toward these systems through a circumbinary disk
because interstellar gas at the center of merged galaxies is expected to have
far too much angular momentum to approach the binary directly
\citep{Springel2005, Chapon2013}. Circumbinary disks differ in many
respects from accretion disks around single black holes (BHs),
especially for mass-ratios close to unity. The most striking differences
originate in the strong gravitational torques that the orbiting BHs
exert on the surrounding matter. Early one-dimensional work suggested
that these torques would prevent any gas from falling toward the binary
\citep{Lin1979, Pringle1991}, but more recent multi-dimensional
simulations showed that most of the externally-supplied mass is accreted
and an approximate inflow equilibrium can be reached \citep[see, for
instance,][]{Artymowicz1996, MacFadyen2008, Noble2012, Shi2012,
DOrazio2013, Farris2014a, Zilhao2015, Shi2015, Rafikov2016, Tang2017, Miranda2017}.

These simulations also showed that the circumbinary disk is
truncated at a distance $\approx 2b$ from the binary center-of-mass,
where $b$ denotes the binary separation.  Outside this truncation radius,
mass piles up, forming a local peak in the surface density profile;
inside this radius, the accretion flow onto the binary is confined
within two narrow streams traversing a low-density gap. Each of these
streams is associated with one of the black holes.

These streams are complex systems.   A portion of their mass
receives enough angular momentum from the binary torques to be flung
back to the inner edge, transferring significant amounts of angular
momentum to the disk.   Their impact can cause a steady growth in
an $m=1$ mode of the azimuthal distribution of matter at the inner
edge, giving rise to an orbiting overdensity at $2b<r<4b$ that can be
the dominant source of matter for accretion onto the binary.  This is
the so-called ``lump".

When the accretion streams enter the binary, each feeds a small
accretion disk surrounding one of the black holes; the so-called \textit{mini-disks}
\citep[see, for instance,][]{Hayasi2007, Farris2012, Farris2014a, Gold2014a, Bowen2017, Bowen2018, Bowen2019,
Munoz2019, Munoz2020, Moody2019}. If the binary separation is more
than a  few tens of $M$, where we use geometrical units and $M$ is the mass of the system,
accretion of each mini-disk is slow because it is limited by internal angular momentum transport.
However, at smaller binary separations, the ratio between the radius of each minidisk's
outer edge and its ISCO shrinks to be only order unity; in that situation,
the need for angular momentum transport diminishes,
and their mass content becomes far more time-variable
\citep{Bowen2017,Bowen2018,Bowen2019}.
Efforts toward producing realistic spectra from these simulations have
 begun \citep{Dascoli2018}.

The techniques used to achieve these results are highly diverse, but a
key feature of the system has remained elusive: the spin of the BHs. In
fact, most of these works assume Newtonian gravity, while black hole
spin is inherently relativistic.  Although some works evolved the full set of
Einstein Field Equations (EFE) for the metric of the spacetime
and matter fields \citep{Farris2011, Bode2011,
Giacomazzo2012, Gold2014b}, they focused on binaries close to merger ($b \leq 10M$),
where the strong tidal interactions and short inspiral timescale prevent
the formation of mini-disks. An intermediate strategy in modeling the
gravitational field has been to employ an approximate metric for the
background spacetime, still capturing the relativistic nature of the
system while being freed from the computational load of integrating the
EFE. For instance, \cite{Noble2012} excised the inner region and used a
Post-Newtonian (PN) metric of order 2.5-PN to integrate the equations of
general relativistic magnetohydrodynamics (GRMHD) in the circumbinary
region. \citet{Bowen2017, Bowen2018, Bowen2019} used the global
approximate metric of \cite{Mundim2014} to explore the relativistic
dynamics of the mini-disks in non-spinning binaries. The approach of
\cite{Mundim2014} for building the approximate spacetime can be
generalized to spinning binaries \citep{Gallouin2012, Ireland2016}, but
the analytical metric becomes too complex and computationally expensive
for GRMHD simulations.

In this work, we perform the first simulations of this system
including spin.  To do so, we construct a new approximate metric for
the spacetime of a pair of spinning black holes by linearly
superimposing two individual BHs in the Kerr-Schild gauge. This new
metric, which we call \textit{Superposed Kerr-Schild} (SKS), presents
many advantages over our previous approach.
It is well-behaved in every region of spacetime, it is
easy to implement, it is computationally efficient, and it permits easy
inclusion of spin. Using this approximate spacetime, we have been
able to conduct lengthy GRMHD simulations of circumbinary accretion onto
a SMBBH whose black holes spin.

Our work is organized as follows. In Section \ref{sec:spacetime} we
introduce the Kerr-Schild gauge and construct the SKS metric for the BBH
spacetime. In Section \ref{sec:spacetimeanalysis} we analyze the validity of the
SKS metric as a solution of EFE in vacuum.
Then, in Section \ref{sec:simulations}, we describe the
configurations of our GRMHD simulations of circumbinary disks around
spinning binaries.
Section \ref{sec:results} is devoted to our main
results on the effects of the spins on the properties of the evolved circumbinary disks.
Finally, in Section \ref{sec:conclusions}, we summarize our main conclusions.
Throughout this paper, Latin indices denote spatial indices, running
from 1 to 3; Greek indices denote spacetime indices, running from 0 to 3
(0 is the time coordinate); and the Einstein summation convention is used.
We work on geometrical units where $G=c=1$, and the total mass of the binary $M$
is normalized to unity.

\section{Spacetime construction}    \label{sec:spacetime}

    \subsection{Kerr-Schild form for single black holes}
    
    The Kerr-Schild form of the metric for the spacetime of a single,
    rotating, black hole is \citep[see the republication][]{KerrSchild2009}
    
    \begin{equation}
    \label{kerrschild}
        \gmunu = \etamunu + 2 \mathcal{H} \lmu \lnu \,,
    \end{equation}
    where, in Kerr-Schild Cartesian coordinates $\xalpha=(t,\,x,\,y,\,z)$, $\etamunu= \mathrm{diag}(-1,\,1,\,1,\,1)$,
   $\lmu$ denotes a null vector with respect to both metrics
    $g^{\mu \nu} \lmu \lnu = \eta^{\mu \nu} \lmu \lnu = 0$, and $\mathcal{H}$
    is a scalar function of coordinates. These read:
    \begin{eqnarray}
    \lmu = \left(1,\, \frac{r_{\mathrm{KS}} x + a y}{r_{\mathrm{KS}}^2 + a^2},\, \frac{r_{\mathrm{KS}} y - a x}{r_{\mathrm{KS}}^2 + a^2},\, \frac{z}{r_{\mathrm{KS}}}\right) \,,
\\
        \mathcal{H} = \frac{M r_{\mathrm{KS}}}{r_{\mathrm{KS}}^2 + a^2 \cos^2 \theta} \,,
    \end{eqnarray}
    with the auxiliary functions
    \begin{eqnarray}
    \label{rKS}
	    {r_{\mathrm{KS}}}^2 = \frac{1}{2} \left( {\rho}^2 - a^2\right) + \sqrt{\frac{1}{4} \left({\rho}^2 - a^2 \right)^2 + {a^2 z^2}} \,,
\\
    \label{rhoKS}
	    {\rho}^2 = x^2 + y^2 + z^2 \,,
\\
	    \cos \theta = \frac{z}{r_{\mathrm{KS}}} \,,
    \end{eqnarray}
    being $M$ the \textit{Arnowitt-Deser-Misner} \citep[ADM; see][]{Baumgarte2010} mass of the black hole, and $a$ its specific angular momentum.

    Kerr-Schild coordinates are widely used for their computational advantages. The coordinates $\xalpha$
    are horizon-penetrating, and singular regions are contained within the event horizon.
    This allows the excision of singularities from the computational domain while keeping the physics at the exterior of the black hole unaffected.
    Furthermore, the Kerr-Schild form is invariant under a Lorentz-boost transformation:
    \begin{eqnarray}
        \label{boost1}
        \xbaralpha = \Lambdaalphabeta \xbeta \,,
\\
        \label{boost2}
	    \bar{\mathcal{H}} \left(\xbaralpha\right) = \mathcal{H}\left({[\Lambda^{-1}]^{\alpha}}_{\beta}\xbarbeta\right) \,,
\\
        \label{boost3}
	    \bar{l}_{\mu} \left(\xbaralpha\right) = {\Lambda^{\nu}}_{\mu}\lnu\left({[\Lambda^{-1}]^{\alpha}}_{\beta}\bar{x}^{\beta}\right) \,,
    \end{eqnarray}
    where $\Lambdaalphabeta$ are the components of the usual Lorentz matrix for uniform velocity $v^{i}$.
    The resulting metric represents the spacetime of a moving, rotating, black hole.

    The invariance of the Kerr-Schild form under boosts is useful for approximating
    the spacetime of multiple moving black holes by linearly superposing terms of the form
    $2 \mathcal{H}^{(n)} \lmu^{(n)} \lnu^{(n)}$ to the same asymptotic background $\eta_{\mu \nu}$, where $n=1,\,2,\,...$ accounts
    for each black hole, with mass $M^{(n)}$ and specific angular momentum $a^{(n)}$.

    \subsection{Superposed Kerr-Schild as initial data for black hole binaries}
    As mentioned in Section~\ref{sec:introduction}, our approach for evolving the spacetime of spinning BBHs is to construct an approximate metric based on the superposition of two Kerr-Schild black holes.
    In this subsection, we briefly review how an equivalent superposition has been used for setting initial data (ID) in Numerical Relativity simulations, i.e. simulations that evolve EFE for the spacetime metric \citep[see][for a review]{Duez_2018}.
    Indeed, in the context of Numerical Relativity, one needs a valid set of ID that solves the constraints of EFE at some Cauchy surface.
    The standard case, where one solves the initial metric for a given distribution of matter,
    presents the problem of having to determine 12 degrees of freedom from just 4 equations \cite{Cook_2000}.
    Some techniques have been proposed for fixing free degrees of freedom in advance,
    while simplifying the constraint equations.

    For instance, the Conformal-Transverse-Traceless (CTT) decomposition
    \citep{YorkCTT1971, BowenYorkSolutions1980} asks for a conformally-related spatial metric
    $\tilde{\gamma}_{ij}$, and the trace and conformal traceless part of the extrinsic curvature,
    before solving the remaining 4 degrees of freedom, contained in the conformal factor $\psi$,
    and potential functions $W^i$. This technique is particularly convenient for conformally
    flat spacetimes, where the resulting equations simplify significantly, but
    BBH spacetimes are not conformally flat \citep{DamourPN2000} and this technique possesses some limitations.
    For instance, it is not possible
    to construct ID for BBHs with spins larger than $\sim 0.93$ from a conformally flat approach \citep{Dain_2002, Lousto_2012}.

   An alternative prescription for the conformal metric $\tilde{\gamma}_{ij}$ was introduced by \cite{MatznerSKS2000}, based on the linear superposition of two boosted Kerr-Schild black holes:
   
    \begin{equation}
        \label{MatznerSKS}
        \tilde{\gamma}_{ij} = \delta_{ij} + 2 \bar{\mathcal{H}}^{(1)} \bar{l}_i^{(1)} \bar{l}_j^{(1)} + 2 \bar{\mathcal{H}}^{(2)} \bar{l}_i^{(2)} \bar{l}_j^{(2)} \,.
    \end{equation}
    Using this approach, \cite{MarronettiID2000a} and \cite{MarronettiID2000b} solved the resulting constraint
    equations for $\psi$ and $W^i$ and found that the solution was in good agreement with the
    conformal ansatz \eqref{MatznerSKS}, even for close separations ($\sim10 M$). Additionally,
   \cite{BonningID2003} supported this claim, and demonstrated that this proposal is well suited for capturing the physics of the BBH
   inspiral as it contains the right Newtonian binding energy for wide separations ($b>15M$).
   More recently, \cite{Lovelace_2008, Lovelace_2012, Scheel_2015, Healy_2016, Ruchlin_2017, Zlochower_2017}
   used the superposition of conformally Kerr-Schild black holes to develop
   new ID that can be used to evolve  BBH with spins as high as $0.994$.
   We conclude that, although some junk gravitational radiation might be present  \citep{PfeifferID2002}, the ansatz \eqref{MatznerSKS} approximates the spacetime of widely separated BHs at a given time.

    \subsection{Time-dependent Superposed Kerr-Schild for binary black hole evolution}
    
    Motivated by the success of the superimposed prescription~\eqref{MatznerSKS} as ID, we model the four-dimensional spacetime of a
    BBH system with a superimposition of two boosted Kerr-Schild black holes, updating the position
    and velocity of each black hole for a given trajectory.
    We call this metric \textit{Superposed Kerr-Schild (SKS)} and reads:
    
    \begin{equation}
    \label{eq:SKSmetric}
	    g_{\mu \nu} = \eta_{\mu \nu} + 2 \hat{\mathcal{H}}^{(1)} \hat{l}^{(1)}_{\mu} \hat{l}^{(1)}_{\nu} + 2 \hat{\mathcal{H}}^{(2)} \hat{l}^{(2)}_{\mu} \hat{l}^{(2)}_{\nu} \,,
    \end{equation}
    where
    \begin{eqnarray}
        \hat{x}^{(n)\alpha} = \Lambda^{(n)}_{\mathrm{circ}}\left(x^{\alpha}\right) \,,
\\
	\hat{\mathcal{H}}^{(n)} (\hat{x}^{(n)\alpha}) = \mathcal{H}^{(n)}\left[{\Lambda^{(n)}_{\mathrm{circ}}}^{-1}\left(\hat{x}^{\alpha}\right)\right] \,,
\\
	    \hat{l}^{(n)}_{\mu} (\hat{x}^{(n)\alpha}) = {\Lambda^{(n)\nu}}_{\mu} l^{(n)}_{\nu}\left[{\Lambda^{(n)}_{\mathrm{circ}}}^{-1}\left(\hat{x}^{\alpha}\right)\right] \,.
    \end{eqnarray}
    We apply standard Lorentz transformations $\Lambda^{(n)}$ to vector fields $l^{(n)}$,
    so we keep an inertial frame $\etamunu$ at infinity, but we apply a non-linear
    transformation $\Lambda^{(n)}_{\mathrm{circ}}$ to coordinates in order to force the BHs to move on
    the desired trajectory. We call this last transformation \textit{circular boost} and we introduce it below.

    Each black hole is boosted with a different velocity $v^{(n)i}$, updated as a function
    of time to be the tangential velocity of a given orbit.
    In a first approximation, we consider equal-mass black holes, and Keplerian circular trajectories in the $x$-$y$ plane,
    with separation $b$:
    \begin{eqnarray}
	    \label{traj1}
	    x_{\rm K}^{(1)} = \frac{b}{2} \cos(\phi)\,, \ \ y_{\rm K}^{(1)} = \frac{b}{2} \sin(\phi)\,, \ \ z_{\rm K}^{(1)} = 0\,,
\\
	    \label{traj2}
	    x_{\rm K}^{(2)} = -\frac{b}{2} \cos(\phi), \ \ y_{\rm K}^{(1)} = -\frac{b}{2} \sin(\phi), \ \ z_{\rm K}^{(2)} = 0 \,,
    \end{eqnarray}
    where $\phi = \Omega t$, and
    \begin{equation}
        \Omega = \sqrt{\frac{M^{(1)}+M^{(2)}}{b^3}} \,.
    \end{equation}
    The time-dependent velocities are derived by $ v^{(n)i}= dx_{\rm K}^{(n)i}/dt$.

    The non-linear transformation $\Lambda^{(n)}_{\mathrm{circ}}$ is constructed as follows:
    A standard boost of coordinates $\Lambda^{(n)}$ results on the BH moving on a straight line,
    with uniform velocity $v^i$. This is encoded on time dependent terms of the form $v^{i} t$
    in the transformation. Here, we replace such terms with the trajectories $\left(x_{\rm K}^{(n)},\,y_{\rm K}^{(n)},\,z_{\rm K}^{(n)}\right)$
    given by Eqs.~\eqref{traj1} and \eqref{traj2}. The transformation reads:
\begin{widetext}
    \begin{equation}
    \label{circularboost}
        {\Lambda^{(n)}_{\mathrm{circ}}}^{-1}\left(\hat{x}^{\alpha}\right) = \left[ \begin{array}{ccccccc}
        
            \gamma^{(n)} \hat{t} - \gamma^{(n)} \check{v}^{(n)x} \hat{x} - \gamma^{(n)} \check{v}^{(n)y}  \hat{y}
        
        \\
        \\
        
            - x_{\rm K}^{(n)} + \left[1 + \left(\gamma^{(n)}-1\right) \left(\check{v}^{(n)x}\right)^2 \right] \hat{x}
            +\left[\left(\gamma^{(n)}-1\right) \check{v}^{(n)x} \check{v}^{(n)y}\right] \hat{y}
        
        \\
        \\
        
            - y_{\rm K}^{(n)} + \left[\left(\gamma^{(n)}-1\right) \check{v}^{(n)y} \check{v}^{(n)x}\right] \hat{x} +\left[1+\left(\gamma^{(n)}-1\right)\left(\check{v}^{(n)y}\right)^2\right] \hat{y}
        
        \\
        \\
        \hat{z}  \end{array} \right] \,,
    \end{equation}
\end{widetext}
    where $\check{v}^{(n)i} = v^{(n)i}/|\vec{v}^{(n)}|$ are the normalized components of the boost velocities.
    The transformation $\Lambda^{(n)}_{\mathrm{circ}}$ is non-linear in the sense that it cannot be written as the linear product of a matrix and the coordinates $\hat{x}^{\mu}$.
    However, its expansion for short intervals of time reduces to a standard boost ${\Lambda^{(n)}}^{-1}$ to the rest frame of the BH, followed by a solid translation to the BH center.

    \section{Spacetime validation}    \label{sec:spacetimeanalysis}

      The SKS metric represents an approximate vacuum solution of EFE.
      Therefore, it should approximately satisfy $R_{\mu \nu} = 0$, where $R_{\mu\nu}$ is the Ricci tensor.
      To quantify deviations from a vacuum solution, following \cite{Mundim2014}, we calculate and analyze the Ricci scalar $R:=R_{\mu \nu}g^{\mu \nu}$ for the SKS metric.

      Even though the metric is analytical,
      we compute the required derivatives numerically. We include
      the SKS metric~\eqref{eq:SKSmetric} in a
      stand-alone code that builds a uniform, three-dimensional, Cartesian grid, and computes the required
      first and second\rm{-order} derivatives by fourth-order finite differences.
      We evaluate the metric components at the corners of the Cartesian cells
      since these positions are shared by different resolutions and this is useful for later
      convergence analysis.

      \begin{figure*}\centering
          \includegraphics[width=.8\linewidth]{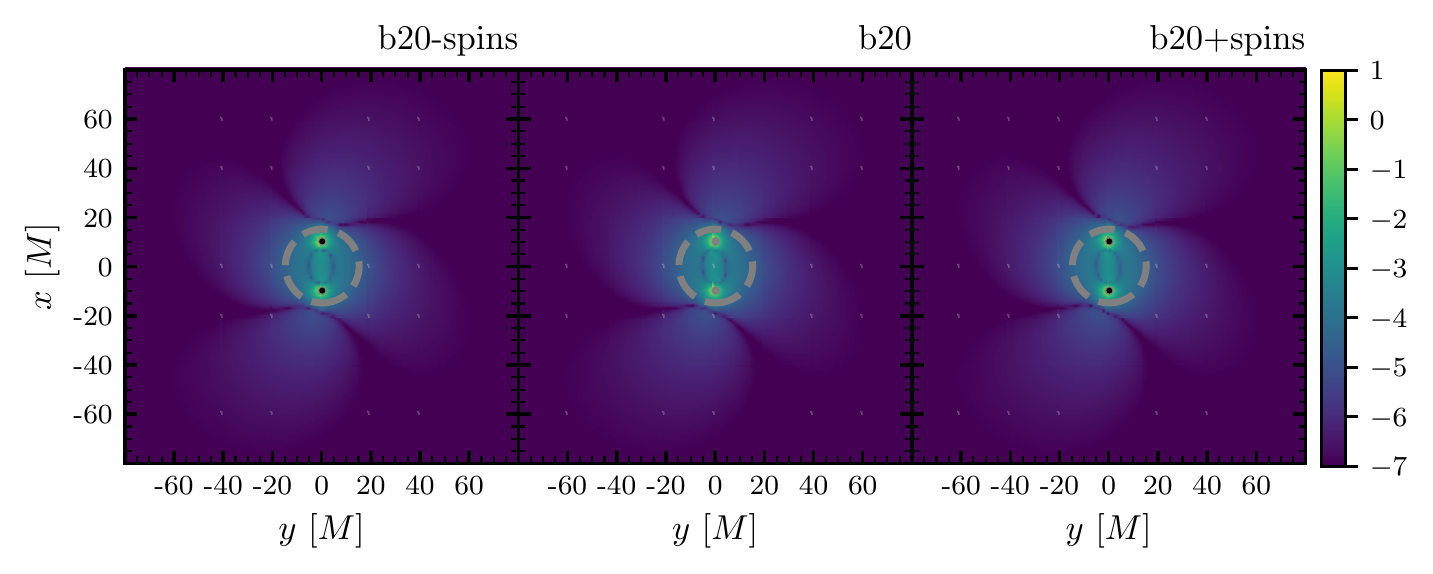}
          \includegraphics[width=.8\linewidth]{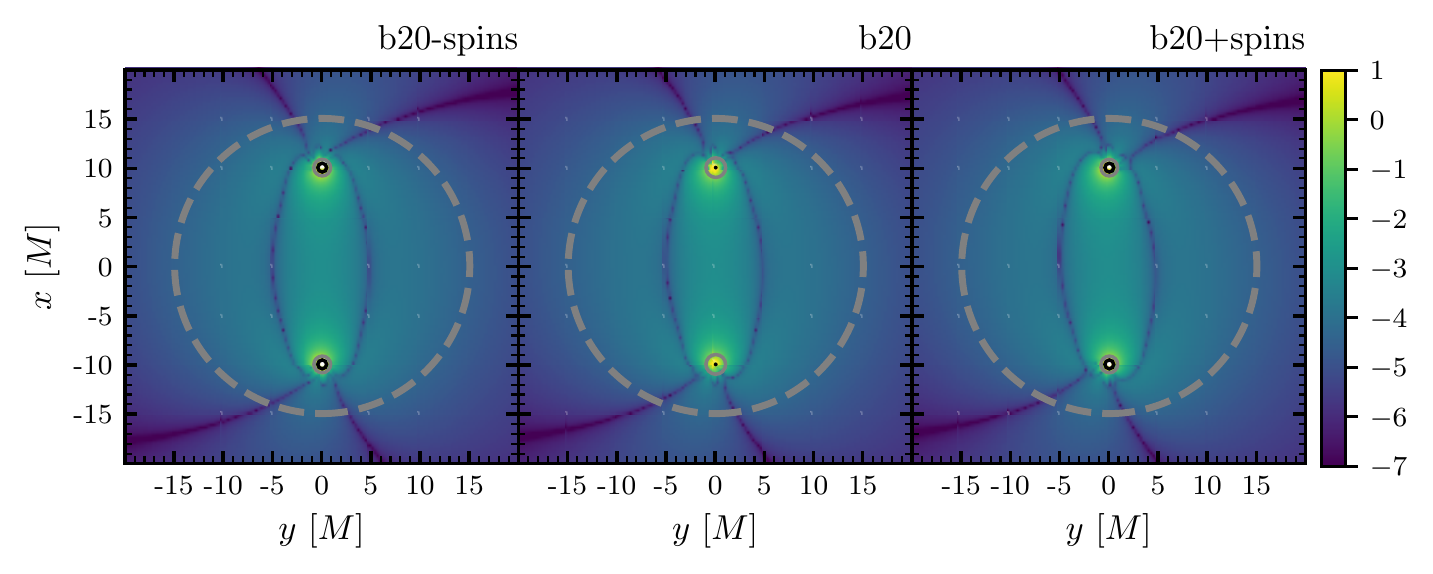}
          \includegraphics[width=.8\linewidth]{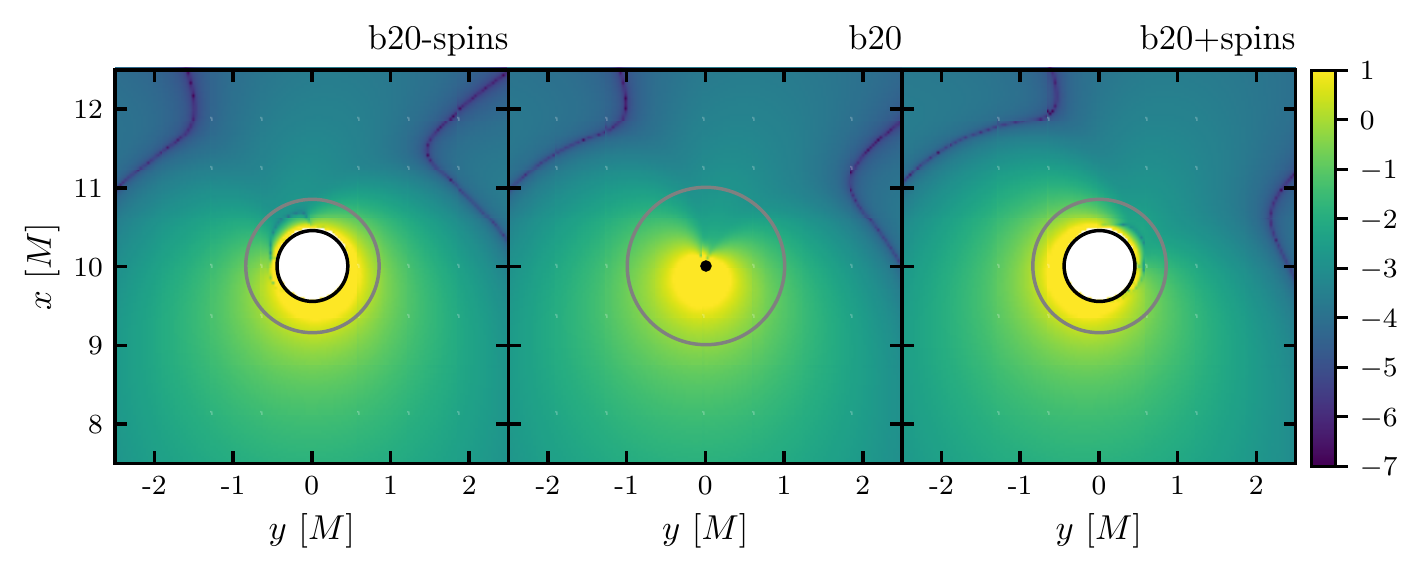}
          \caption{Ricci scalar at the equator for the SKS metric for anti-aligned spins \textit{(left)},
      	non-spinning \textit{(center)}, and aligned spins \textit{(right)}, at different scales
      	\textit{(top, center, bottom)}. Solid-grey circles estimate the BHs horizons at
      	$r^{(1,2)2}=2M^{(1,2)}\left( M^{(1,2)} + \sqrt{ {M^{(1,2)}}^2 - {a^{(1,2)}}^2}\right)$,
      	solid-black circles estimate the BHs singular regions at $r^{(1,2)}=\left|a^{(1,2)}\right|$, where $r^{(n)}$ is the radial distance from the $n$-th BH.
        These estimations follow from known singularities and horizons for a single BH in Cartesian-KS coordinates.
      	Dashed circles represent the limit of the excised region of the domain of our GRMHD simulations, at $r=15M$.
      	The Ricci scalar is calculated through 4th-order finite differencing of the SKS metric~\eqref{eq:SKSmetric},
 in Cartesian-KS coordinates, with $320 \times 320 \times 80$ cells,
      	for grid lengths of $(160,\,160,\,40)$ \textit{(top)}, $(40,\,40,\,10)$ \textit{(center)}, and $(5,\,5,\,1.25)$
      	\textit{(bottom)}, respectively for $x,y,z$.}
        \label{f:Ricci}
      \end{figure*}

      In Fig.~\ref{f:Ricci}, we plot the resulting values of the Ricci scalar $R$
      at $t=0$ in the plane of the BHs, for different spatial scales, and spin values.
      We study the case of equal-mass BHs $M^{(1,2)}=0.5M$, separated by $b=20M$.
      The top row shows the results for a grid centered at the center of mass of the system.
      For coordinates $(x,y,z)$ the domain dimensions are $(160,\,160,\,40)\, M$, and there are $320 \times 320 \times 80$ cells.
      In the left, middle, and right columns, the BH spins are $-0.9 M^{(1,2)}$, $0$ and $0.9 M^{(1,2)}$, respectively.
      Within the circumbinary region, the violations of $R=0$ are comparable to those of \cite{Mundim2014}, and this result is independent of spin for the three cases we explored.

      In the middle row of Fig.~\ref{f:Ricci}, we focus on the orbital region
      by reducing the grid dimensions  to $(40,\,40,\,10) \: M$ while keeping the same number of cells.
      In the domain of the binary ($r<b/2$), particularly between the BHs, the quality of the spacetime
       is not as good as in the circumbinary region, but the values of $R$ are still comparable to those of \cite{Mundim2014}.
      The bottom row shows the Ricci scalar
      $R$ for grid lengths of $(5,\,5,\,1.25) \: M$, centered on one of the BHs, while keeping the same number of cells.
      We notice the metric captures the singularities of spinning BHs.

\begin{figure}
    \centering
    	\includegraphics[width=.8\linewidth]{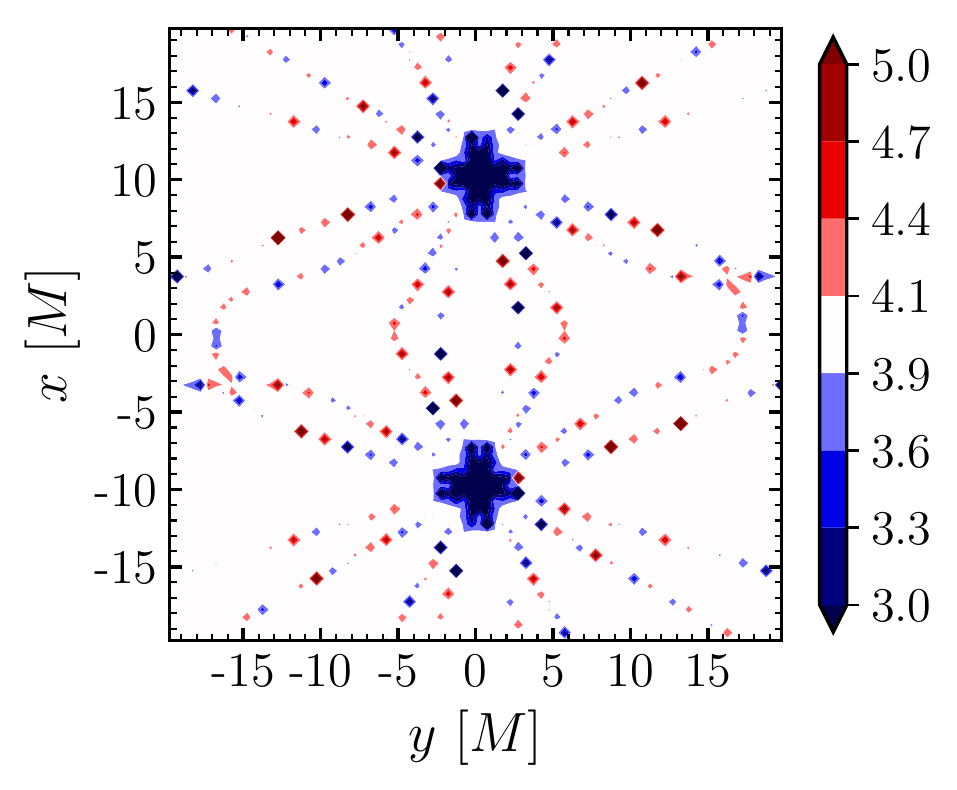}
    	\caption{Convergence factor $p_R$ (Eq.~\eqref{eq:pR}) in linear colorscale, measured in the BH orbital plane, as determined from    	resolutions: $\Delta_1 x=0.5,\, \Delta_2x=0.25,\, \Delta_3x=0.125$.}
      \label{f:RicciConvergence}
    \end{figure}

      Convergence testing proves that the numerical calculation of the
      Ricci scalar $R$ converges to the analytical value. To that end, we recalculate the Ricci scalar for
      the region of the middle row of Fig.~\ref{f:Ricci} with successively coarser resolutions:
      $160 \times 160 \times 40$ and $80 \times 80 \times 20$.
      Then we calculate the local convergence factor,
      
      \begin{equation}
        \label{eq:pR}
        p_R=\frac{1}{\log 2} \log\left| \frac{R_{\Delta_0}-R_{\Delta_1}}{R_{\Delta_1}-R_{\Delta_2}} \right| \,,
      \end{equation}
      where $R_{\Delta_0}, \, R_{\Delta_1}$ and $R_{\Delta_2}$ are, respectively, the values
      of $R$ for the coarsest, middle, and finest resolutions, computed
      at the corners of the coarsest grid cells because these positions are shared by the three resolutions.
      Fig.~\ref{f:RicciConvergence} shows these values of $p_R$ in the BH orbital plane; throughout this region
      $p_R \approx 4$, as expected from a fourth-order finite differencing
      scheme. The apparent non-convergence in the regions close to the BHs is because the coarsest grid fails to
      resolve such high curvatures.

    In this section we analyzed the Ricci scalar $R$ for the SKS metric and find it is approximately zero, as expected for a vacuum solution of EFE.
    The accuracy of this four-dimensional scalar allows us to use the SKS metric as a time-dependent geometry for the background spacetime in GRMHD simulations. Though not included in this article, we also checked for the Hamiltonian and momentum constraints, and found them to be satisfied to the same degree of accuracy as the Ricci scalar $R$.
    As a further validation test, in Appendix~\ref{s:postnewtonian} we prove the expansion of this metric agrees with the lowest PN expansion of the metric of spinning binaries.

\section{Circumbinary disk models}
\label{sec:simulations}

	As a first application of the SKS metric~\eqref{eq:SKSmetric}, we build and evolve a torus of gas
	in the circumbinary region.
	We evolve the system integrating the GRMHD equations of motion (EoM) with the well-tested code \harm
	\citep{Gammie2003, Noble2006, Noble2009}. We neglect the contribution of matter fields
	to spacetime curvature and use the SKS metric as the background geometry.
	Since we focus on the features of the circumbinary disk, we excise a spherical region at the center of the domain that contains the BHs.

	\subsection{GRMHD evolution}

	The evolution of the circumbinary disk follows from the integration of the GRMHD EoM on the background	 SKS metric.
These equations are the continuity equation, the local conservation of energy and momentum, and Maxwell's equations
\cite[see][]{Noble2009}. In flux-conservative form, these read:

\begin{equation}
        \label{consEqs}
        \partial_t \mathbf{U}(\mathbf{P}) = - \partial_i \mathbf{F}^{i} + \mathbf{S} (\mathbf{P}) \,,
\end{equation}
where $\mathbf{P}$ is the vector of \textit{primitive} variables, $\mathbf{U}$ the vector of \textit{conserved} variables,
$\mathbf{F}$ the \textit{fluxes}, and $\mathbf{S}$ the \textit{sources}. They read:

\begin{equation}
\mathbf{P} = \left[\rho, P, \tilde{u}^k, B^k\right]^{\mathrm{T}},
\end{equation}
\begin{equation}
\mathbf{U}(\mathbf{P}) = \sqrt{-g} \left[\rho u^t, {T^t}_t + \rho u^t, {T^t}_j, B^k\right]^{\mathrm{T}},
\end{equation}
\begin{equation}
\mathbf{F}^i(\mathbf{P}) = \sqrt{-g} \left[ \rho u^i, {T^i}_t + \rho u^{i},\right. \left.{T^i}_j, \left(b^i u^k - b^k u^i \right) \right]^{\mathrm{T}},
\end{equation}
\begin{equation}
        \mathbf{S}(\mathbf{P}) = \sqrt{-g} \left[ 0, {T^{\kappa}}_{\lambda} {\Gamma^{\lambda}}_{t\kappa} - \mathcal{F}_t, \right. \left. {T^{\kappa}}_{\lambda} {\Gamma^{\lambda}}_{j\kappa} - \mathcal{F}_j, 0^k \right]^{\mathrm{T}},
\end{equation}
where $g$ denotes the determinant of the SKS metric, $\rho$ is the rest mass density,
$u^{\mu}$ is the fluid four-velocity, and $\tilde{u}^{\mu}$ is the fluid four-velocity as measured by a zero angular momentum observer (ZAMO).
The magnetic field is represented by $B^k ={}^*F^{kt} \sqrt{4\pi}$, where
${}^*F^{\mu \nu}$ is the dual of the Maxwell tensor,
$b^{\mu} = \left(\delta^{\mu}_{\nu} + u^{\mu} u_{\nu} \right) B^{\nu}$ is the projection of the magnetic field into the fluid's comoving frame.  In addition, ${\Gamma^{\lambda}}_{\mu \nu}$
is the affine connection for the SKS metric and  ${T^{\mu}}_{\nu}$ is the sum of the
stress-energy tensor of a perfect fluid and the EM stress energy tensor, defined as:

\begin{equation}
        T_{\mu \nu} = \left(\rho h + 2 \pmag \right) u_{\mu} u_{\nu} + \left(p + \pmag\right) g_{\mu \nu} - b_{\mu} b_{\nu} \,,
\end{equation}
where $p$ denotes the the pressure of the fluid, $h = 1 + \epsilon + p / \rho$ the specific enthalpy,
$\epsilon$ the specific internal energy, and $\pmag = b^{\mu} b_{\mu} / 2$ the magnetic pressure.
The internal energy is $u=\rho \epsilon$, and
we assume an adiabatic $\Gamma$-law equation of state:
$P=(\Gamma - 1) u$, with $\Gamma = 5/3$, corresponding to a non-relativistic fluid without internal degrees of freedom.

In accretion disks, dissipation converts magnetic and kinetic turbulence into heat. To regulate
the consequent growth in the temperature, we follow \cite{Noble2009} and include a
sink term $\mathcal{F}^{\mu}$ in the conservation equations~\eqref{consEqs}.
This term portrays the effect of optically-thin radiative cooling.
For isotropic emission in the fluid's frame, the sink term takes the form $\mathcal{F}_{\nu} = \Lcool u_{\nu}$,
where $\Lcool$ is the \textit{cooling function}, defined as the rate of radiated energy per unit
of proper time. We set  $\Lcool$ to follow the local increase in entropy $S$, cooling the plasma to the initial entropy $S_0$ \citep{Noble2012}:

\begin{equation}
	\Lcool = \frac{\rho \epsilon}{t_{\mathrm{cool}}} \left( \frac{\Delta S}{S_0} + \left| \frac{\Delta S}{S_0} \right| \right)^{1/2} \,,
	\label{e:Lcool}
\end{equation}
where $\Delta S = S - S_0$, $S=p/\rho^{\Gamma}$, $S_0 = 0.01$, and $t_{\mathrm{cool}} = 2\pi (r/M)^{3/2}$.
We do not cool unbound material,
defined as the portion of the plasma where $\left(\rho h + 2 \pmag\right) u_t < -\rho$.
By controlling the temperature, the cooling function stabilizes the aspect ratio $H/r$ of the disk
(see Eq.~\eqref{e:height}); at the same time, it also estimates the luminosity.
\cite{Noble2012} did not claim to include the square root in Eq.~\eqref{e:Lcool}, but this is a typo in the manuscript and the square root was actually included in the computational code (personal communication).

	We integrate  the conservation equations \eqref{consEqs} with high-resolution shock-capturing
schemes implemented in \harm \citep{Gammie2003, Noble2006, Noble2009}. After reconstruction of the primitive variables to the
cell interfaces through a piecewise parabolic method, we apply the Lax-Friedrichs formula to compute
the local fluxes \citep{Gammie2003}. We use fourth-order finite differences for spatial derivatives, and the
method of lines for time integration with a Runge-Kutta method of second-order. If the updates of $\rho$ or $u$ go
below the corresponding atmosphere values $\rho_{\mathrm{atm}} = 2 \times 10^{-10} (r/M)^{-3/2}$, $u_{\mathrm{atm}} = 2 \times 10^{-12} (r/M)^{-5/2}$
 they are reset to the latter.
The primitive variables are recovered from the conserved variables with the scheme described in \cite{Noble2006}.
We use the constrained transport (FluxCT) algorithm \citep{Toth2000} to maintain the solenoidal constraint, 
$\partial_i \left( \sqrt{-g} B^i \right)=0$.
For more details on the numerical implementation, see \cite{Noble2009}.

 \subsection{Circumbinary disk initialization}

 As initial data for the matter fields, we construct a torus in nearly
hydrostatic equilibrium at the circumbinary region. \cite{Fishbone1976} and \cite{Chakrabarti1985} presented
this solution to the relativistic Euler's equations for the case of stationary and
axisymmetric spacetimes in Boyer-Lindquist (BL) coordinates,
where the only non-zero off-diagonal components of the metric are
$g^{\mathrm{BL}}_{t \phi}$ and $\ g^{\mathrm{BL}}_{\phi t}$.
To use the same technique and build the torus on the stationary and axisymmetric spacetime, we transform the SKS metric~\eqref{eq:SKSmetric} to BL-like coordinates and take a $\phi$-average of the metric as we explain below.

  First, we transform the whole SKS metric~\eqref{eq:SKSmetric}
from Cartesian-KS to BL-like coordinates using the standard
transformations for a single, non-spinning black hole with mass $M = M^{(1)} + M^{(2)}$. The transformation is given by:
\begin{eqnarray} \label{KS2BL}
		 t = u_{\mathrm{BL}} - r_{\mathrm{BL}} + \int dr_{\mathrm{BL}} \frac{r_{\mathrm{BL}}}{r_{\mathrm{BL}} - 2M} \,, \\
		 x_{\mathrm{KS}} = r_{\mathrm{BL}} \sin{\theta_{\mathrm{BL}}} \cos{\phi_{\mathrm{BL}}} \,, \\
		 y_{\mathrm{KS}} = r_{\mathrm{BL}} \sin{\theta_{\mathrm{BL}}} \sin{\phi_{\mathrm{BL}}} \,, \\
		 z_{\mathrm{KS}} = r_{\mathrm{BL}} \cos{\theta_{\mathrm{BL}}} \, \end{eqnarray}
				 
The metric transforms in the usual way:

\begin{equation}
g_{\mu \nu}^{\mathrm{BL}} = \frac{dx^{\alpha}_{\mathrm{KS}}}{dx^{\mu}_{\mathrm{BL}}}
\frac{dx^{\beta}_{\mathrm{KS}}}{dx^{\nu}_{\mathrm{BL}}} g_{\alpha \beta}^{\mathrm{KS}} \,.
\end{equation}

 The SKS spacetime has a helical Killing symmetry through the Killing vector
$\mathcal{K}^{\mu} =  \left(\partial_t\right)^{\mu} + \Omega_{\mathrm{bin}} \left(\partial_{\phi} \right)^{\mu}$, where $\Omega_{\mathrm{bin}}$ is the binary orbital frequency,
and the time average of the metric coincides with the corresponding azimuthal average.
Then, following the procedure of \cite{Noble2012}, we construct a stationary and axisymmetric
spacetime from the azimuthal average:

 \begin{equation}
	 \label{eq:SKSphiav}
	 \tilde{g}^{\mathrm{BL}}_{\mu \nu} = \frac{\int g^{\mathrm{BL}}_{\mu \nu} \sqrt{g^{\mathrm{BL}}_{\phi \phi}} \ d\phi}{\int \sqrt{g^{\mathrm{BL}}_{\phi \phi}} \ d\phi} \,.
 \end{equation}

We then follow the steps of \cite{Noble2012} for the construction of the torus over the metric \eqref{eq:SKSphiav}. The free parameters of this model
are the radial distance to the disk inner edge $r_{\mathrm{in}}$, the radial distance
to the maximum of pressure $r_{\mathrm p}$, and the
specific angular momentum of the fluid at the inner edge $l_{\mathrm{in}}$. From such a procedure, we obtain the
hydrodynamic properties of the fluid, including its four-velocity in BL coordinates $u^{\mu}_{\mathrm{BL}}$.
Transforming the four-velocity to Cartesian-KS via $u^{\mu}_{\mathrm{KS}} =
\frac{dx^{\mu}_{\mathrm{KS}}}{dx^{\alpha}_{\mathrm{BL}}} u^{\alpha}_{\mathrm{BL}}$, we obtain the hydrodynamical ID
for the torus in a coordinate system consistent with the SKS metric~\eqref{eq:SKSmetric}.
We include random perturbations of the internal energy $u=\rho \epsilon$, with amplitude $10^{-2}$,
to precipitate turbulence and accretion.

We initialize the magnetic field in the interior of the disk as
a set of dipolar loops that follow the lines of constant density of the fluid. The corresponding
vector potential $A_{\mu}$ in spherical coordinates $(t,\,r,\,\theta,\,\phi)$ has one non-vanishing component:

\begin{equation}
	A_{\phi} = A_0 \mathrm{max} \left[\left( \rho - \rho_{\mathrm{cut}} \right),\, 0 \right] \,,
 \end{equation}
where $\rho_{\mathrm{cut}} = 0.25\, \rho_{\mathrm{max}}$ so the initial magnetic field is
entirely confined within the torus and the field lines wrap around the region of maximum density $\rho_{\mathrm{max}}$.
The constant $A_0$
is chosen such that the initial ratio of the fluid integrated pressure to the magnetic integrated pressure satisfies:

\begin{equation}
\frac{\int p \sqrt{-g}\, d^{3}x}
{\int p_{\mathrm{m}}\, \sqrt{-g}\, d^{3}x} \sim 100.
\label{e:ratioP}
\end{equation}
In this way, the initial equilibrium between thermal and magnetic stresses
is comparable for different simulations. In the next subsection, we explain how these
spherical coordinates $(t,\,r,\,\theta,\,\phi)$ relate to the Cartesian-KS.
\cite{Noble2012} claimed that the ratio of the fluid's total internal energy to the total magnetic energy was initialized to $100$, but this is a typo in the manuscript and actually the condition~\eqref{e:ratioP} was demand (personal communication).

 We fill the region outside the torus with an atmosphere, or numerical vacuum, modeled
by a tenuous, non-magnetized, static fluid in approximate
hydrostatic equilibrium:
$\rho_{\mathrm{atm}} = 2 \times 10^{-10} (r/M)^{-3/2}$, $u_{\mathrm{atm}} = 2 \times 10^{-12} (r/M)^{-5/2}$,
 and $u_{\mathrm{atm}}^i = 0$.

 The last step of the initialization involves the transformation of the SKS metric~\eqref{eq:SKSmetric},
 the initial four-velocity $u^{\mu}_{\mathrm{KS}}$, and the initial four-vector potential $A_{\mu}$, to the numerical
 coordinates used for the integration of the GRMHD EoM (see the next subsection).

\subsection{Numerical grid, boundary conditions, and simulation parameters}
\label{s:sim_params}

    	Numerical errors in conservation of momentum are smallest in the direction of coordinate
	 lines; consequently, given the approximate axisymmetry of the system, the Cartesian-KS basis
	 would be a poor choice for the global coordinates of the simulation. We move to a spherical basis through
	 a standard spatial transformation:
\begin{eqnarray}
 t = t \ , \\
 x_{\mathrm{KS}} = r \sin{\theta} \cos{\phi} \ , \\
 y_{\mathrm{KS}} = r \sin{\theta} \sin{\phi} \ , \\
 z_{\mathrm{KS}} = r \cos{\theta} \ . 
\end{eqnarray}
 These are the physical coordinates in our simulation~\footnote{These coordinates are not the usual spherical Kerr-Schild coordinates used in the literature of accretion disks (see, for instance, \cite{Gammie2003}), but they result from a standard spherical transformation of the Cartesian Kerr-Schild coordinates usually used in the literature of Numerical Relativity (see, for instance, \cite{MatznerSKS2000})}.

	 For the actual integration of the EoM
	 we move to a numerical coordinate system $\left( x^{(0)},\,x^{(1)},\,x^{(2)},\, x^{(3)} \right)$
	 that relates to the physical one by:
	 \begin{eqnarray}
	 			t = x^{(0)}  \\
	 	r\left(x^{(1)}\right) = M e^{x^{(1)}}  \\
	 	\theta\left(x^{(2)}\right) = \frac{\pi}{2} \left[1 + (1-\xi)\left(2 x^{(2)} -1 \right) + \right.  \\  \nonumber + \left. \left(\xi - \frac{2\theta_c}{\pi}\right)\left(2 x^{(2)} - 1\right)^n \right]   \\
	 	\phi\left(x^{(3)}\right)  = x^{(3)} \,,
	 \end{eqnarray}
	 where $n=9$, $\xi=0.87$, and $\theta_c = 0.2$. We construct a uniform grid of $\left(x^{(1)},\,x^{(2)},\,x^{(3)}\right)$
where the center of the $i,\,j,\,k$-cell has coordinates $\left(x^{(1)}_i,\,x^{(2)}_j,\,x^{(3)}_k\right)$, with
$x^{(1)}_i = x_{\mathrm{b}}^{(1)} + (i+1/2) \Delta x^{(n)}$, and equivalently for $x^{(2)}_j$ and $x^{(3)}_k$.
The grid, then, is determined by the parameters: $x_{\mathrm{b}}^{(1)} = \ln(r_{\mathrm{min}}/M)$,
$\Delta x^{(1)} = \ln(r_{\mathrm{max}}/r_{\mathrm{min}})/N^{(1)}$, $r_{\mathrm{min}}=15 M$,
$r_{\mathrm{max}} = 300 M$, $N^{(1)}=300$, $x_{\mathrm{b}}^{(2)} = 0$,
$\Delta x^{(2)} = 1/N^{(2)}$, $N^{(2)} = 160$, $x_{\mathrm{b}}^{(3)} = 0$, $\Delta x^{(3)} = 2 \pi / N^{(3)}$, and $N^{(3)}=400$.
A uniform grid of these numerical coordinates implies better resolution at smaller physical radii and at the equatorial plane of the system.
\cite{Noble2012} showed that, in these conditions, this grid resolves the magnetorotational instability (MRI)~\citep{velikhov1959_MRI,chandrasekhar1960_MRI,BalbusHawley1991_MRI} and spiral density waves generated by the binary torques~\footnote{There is a small difference between our grid and the one used by \cite{Noble2012}. The latter set  $r_{\mathrm{max}}=260M$, but we extend it to $r_{\mathrm{max}} = 300M$. Our grid still satisfies the
physical resolution requirements.}.
We evolve this sytem from $t=0$ to $t=1.5 \times 10^5 M$, using a
dynamical step $\Delta t = 0.45 \Delta t_{\mathrm{min}}$, where $\Delta t_{\mathrm{min}}$ is
the shortest cell crossing time of matter fields over the domain at each time.

  Boundary conditions are imposed through zeroth-order extrapolation of primitive variables
into ghost zones. Specifically, outflow boundary conditions are applied on $x^{(1)}$ and $x^{(2)}$-boundaries,
while periodic boundary conditions are used on $x^{(3)}$-boundaries. We force $u^r=0$ if it points into the domain at the $r$-boundaries.
This diode-type condition was found to be unstable in some circumstances involving low-density regions by \cite{Noble2012} but successfully
used in Newtonian simulations by \cite{MacFadyen2008}, \cite{Shi2012}, \cite{DOrazio2013}, among others.

	We perform a set of five runs, denoted: \runmspins, \runvz , \runvo, \runvt, \runpspins.
In every run, the BHs have equal masses: $M^{(1)} = M^{(2)} = 0.5$,
so the total mass of the system is $M=1$; the distance between them is fixed to $b=20M$.
The disk's initial inner edge is at $r_{\mathrm{in}}=60M$,
and the initial pressure maximum is at $r_{\mathrm p} = 100M$.

  The spins of the BHs in \runmspins{} are $a^{(1)}=a^{(2)}=-0.9 M^{(1,2)}$, i.e. opposite to the angular momentum of the binary. The spins in run \runpspins{} have the same magnitude but are aligned with the orbital angular momentum.  Runs \runvz{}, \runvo{}, \runvt{} have no spin.   These three runs differ from one another only in the random initial perturbations of the internal energy; the goal of these runs is to calibrate the size of intrinsic fluctuations due to turbulence so that we can tell whether the spin runs differ significantly.  In Table \ref{t:runs} we gather the relevant  properties of the binaries and initial disks of our runs.

  The specific angular momentum of the fluid at the inner edge of the disk $l_{\mathrm{in}}$
is set so the ratio $H/r$ equals $0.1$ at $r_{\mathrm p}$. This results in
$l_{\mathrm{in}} = 8.62M$, $8.60M$, $8.60M$, $8.60M$ and $8.57M$ for
\runmspins, \runvz , \runvo, \runvt \ and \runpspins, respectively (see Table \ref{t:runs}).

\begin{table*}
  \begin{center}
    \begin{tabular}{l|c|c|c|c|c|c} 
			$\,$ & $b \, [M]$ & $M^{(1,2)}\, [M]$ & $\Omega_{\mathrm{bin}}$ & $a^{(1,2)}$ & $l_{\mathrm{in}} \, [M]$ & $\Sigma_0 \, [M^{-1}]$ \\
			\hline
			\textbf{\runmspins} & $20 $ & $0.5$  & $b^{-3/2}$ & $-0.9M^{(1,2)}$ & $8.62$ & $0.1070$ \\
			\textbf{\runvz} &  $20 $ & $0.5$ &  $b^{-3/2}$ & $0.0$ & $8.60$ & $0.1066$ \\
			\textbf{\runvo} &  $20 $ & $0.5$ &  $b^{-3/2}$ & $0.0$ & $8.60$ & $0.1066$ \\
			\textbf{\runvt} &  $20 $  & $0.5$ &  $b^{-3/2}$ & $0.0$ & $8.60$ & $0.1066$ \\
			\textbf{\runpspins} & $20 $ & $0.5$ & $b^{-3/2}$ & $0.9M^{(1,2)}$ & $8.57$ & $0.1063$ \\
    \end{tabular}
  \end{center}
	\caption{Properties of the binary system for our runs, and the initial values of $l_{\mathrm{in}}$ and $\Sigma_0$. In every case the BHs separation is fixed to $b=20M$, they have equal masses $M^{(1,2)}=0.5M$, and move in Keplerian orbits with $\Omega_{\mathrm{bin}}=b^{-3/2}$.
	We explore different values for the spins of the BHs.
	Notice runs \runvz, \runvo{} and \runvt{} have identical settings. They only differ on the random initial perturbations on the internal energy $u$.}
	\label{t:runs}
\end{table*}

\section{Circumbinary Disk Dynamics}
\label{sec:results}

\begin{figure*}\centering
        \includegraphics[width=.9\linewidth]{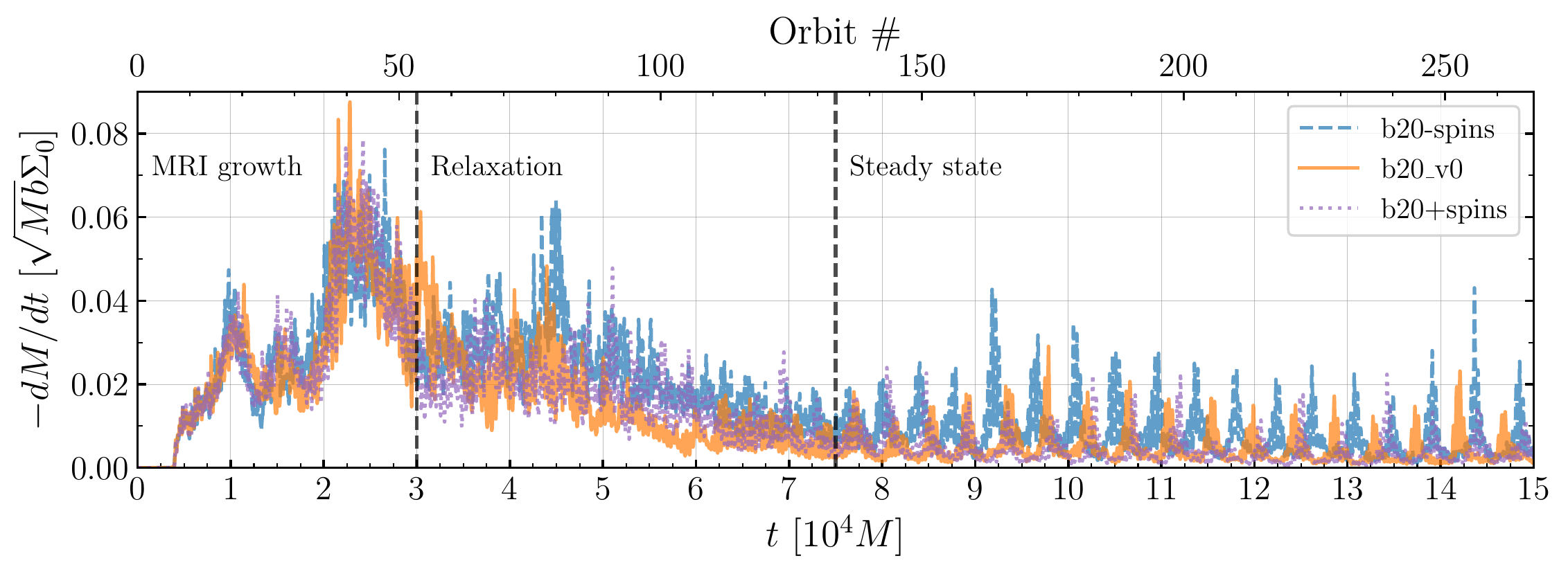}
        \caption{Accretion rate integrated at the innermost boundary of the grid, as a function of time.
        From this plot, we distinguish three dynamical stages: \textit{MRI growth} ($t=0$ -- $30\e{3} M$), \textit{relaxation} ($t=30$ -- $75\e{3} M$), and \textit{steady state} ($t=75$ -- $150\e{3} M$).}
        \label{f:Mdot}
\end{figure*}

To globally characterize the dynamics of these simulations, in Fig. \ref{f:Mdot} we plot the accretion rate $\dot{M}$ as a function of time (see Eq.~\eqref{e:mdot})  at the innermost radial boundary of the domain.
We distinguish three dynamical stages in this plot: MRI growth ($t=0$ -- $30\e{3} M$), in which the MRI grows to its saturated amplitude; subsequent relaxation ($t=30$ -- $75\e{3} M$), in which the accretion rate progressively diminishes over time; and a steady state ($t=75$ -- $150\e{3} M$).
The first is a transient period, and will not be included in our analysis.
The second stage is still affected by the initial transient and will not be used for our main conclusions.
We will focus, instead, on the steady state epoch.

We organize our results in three subsections.
First, we focus on the properties of the plasma that are sensitive to the spins; these properties are mostly related to the cavity and the accretion streams. Then, in the second subsection, we interpret these spin-sensitive results in terms of the gravitational potential of the linearized SKS metric. Finally, in the third subsection, we describe the bulk properties of the circumbinary disk, all of them insensitive to the spin of the BHs.

Because MHD turbulence is a fundamental property of accretion disks, all our results are subject to intrinsic variance.
This fact complicates the identification of subtle physical processes such as the effect of the spins on the circumbinary disk.
To quantify this variance, we use the subset of runs \runvz, \runvo, and \runvt{}.
The parameters of these three runs are identical and their only differences arise from stochastic processes triggered by random initial perturbations in the internal energy of the fluid.
Specifically, given a physical quantity $\mathcal{P}_i$ with $i=0,1,2$ for runs \runvz, \runvo, and \runvt{}, we will express the result as $\left< \mathcal{P} \right>_{a=0}  \pm \sigma_{\mathcal{P}}$, where

\begin{equation}
  \left< \mathcal{P} \right>_{a=0} = \frac{1}{3}\sum_{i=0}^2 \mathcal{P}_i
\end{equation}
is the mean of $\mathcal{P}_i$ over the non-spinning runs, and

\begin{equation}
  \sigma_{\mathcal{P}} = \sqrt{\sum_{i=0}^2 \frac{\left( \left< \mathcal{P} \right>_{a=0} - \mathcal{P}_i \right)^2}{3-1}}
\end{equation}
is a coarse measure of the corresponding standard deviation.
To determine whether a run with different parameters differs significantly from the three non-spinning runs, we measure the deviation $Z$ of its prediction $\mathcal{P'}$, in units of standard deviations by

\begin{equation}
  Z = \frac{ \mathcal{P'} - {\left< \mathcal{P} \right>_{a=0}}}{\sigma_{\mathcal{P}}} \,.
\end{equation}

Following \cite{Noble2012}, many of our results will be expressed in units of $\Sigma_0$, the initial maximum value of the surface density $\Sigma(r,\phi)$ (see Eq.~\eqref{e:Sigma_rphi}).
These values are $\Sigma_0=0.1070M^{-1}, \, 0.1066M^{-1}, \, 0.1066M^{-1}, \, 0.1066M^{-1}$ and $0.1063M^{-1}$ for runs \runmspins{}, \runvz{}, \runvo{}, \runvt{} and \runpspins{}, respectively (see Table~\ref{t:runs}).

\subsection{Spin-Sensitive Results}
\label{sec:spin-sens-results}

The spin of a BH has important effects on matter orbiting near the horizon, but these effects decline rapidly with radius; frame-dragging terms in the effective gravitational potential for spinning black holes are $\propto r^{-3}$ (see Appendix \ref{s:postnewtonian}).
For this reason, we do not expect the spin of the BHs will have a direct impact on the bulk properties of the circumbinary disk, whose inner edge lies at $r \approx 50M$.
The accretion streams, on the contrary, reach distances close enough to the black hole  that these effects may be relevant.

Since the accretion streams carry nearly all the matter accreted by the binary, we begin by exploring the effect of the spins on the accretion rate.
For all three non-spinning cases, the time-averaged accretion rate at the inner boundary during the steady state period is (see Fig.~\ref{f:Mdot}) $(5.0 \pm 0.4) \e{-3} \, \sqrt{M b}\, \Sigma_0$.
 Strikingly, runs \runmspins{} and \runpspins{} deviate from this mean value by $+5.7$ and $-1.8$ standard deviations, respectively.
In other words, the circumbinary accretion rate is enhanced (reduced) by $+45\%$ ($-14\%$) if the spin of the BHs are anti-parallel (parallel) to the angular momentum of the binary.

\begin{figure}\centering
  \includegraphics[width=.9\linewidth]{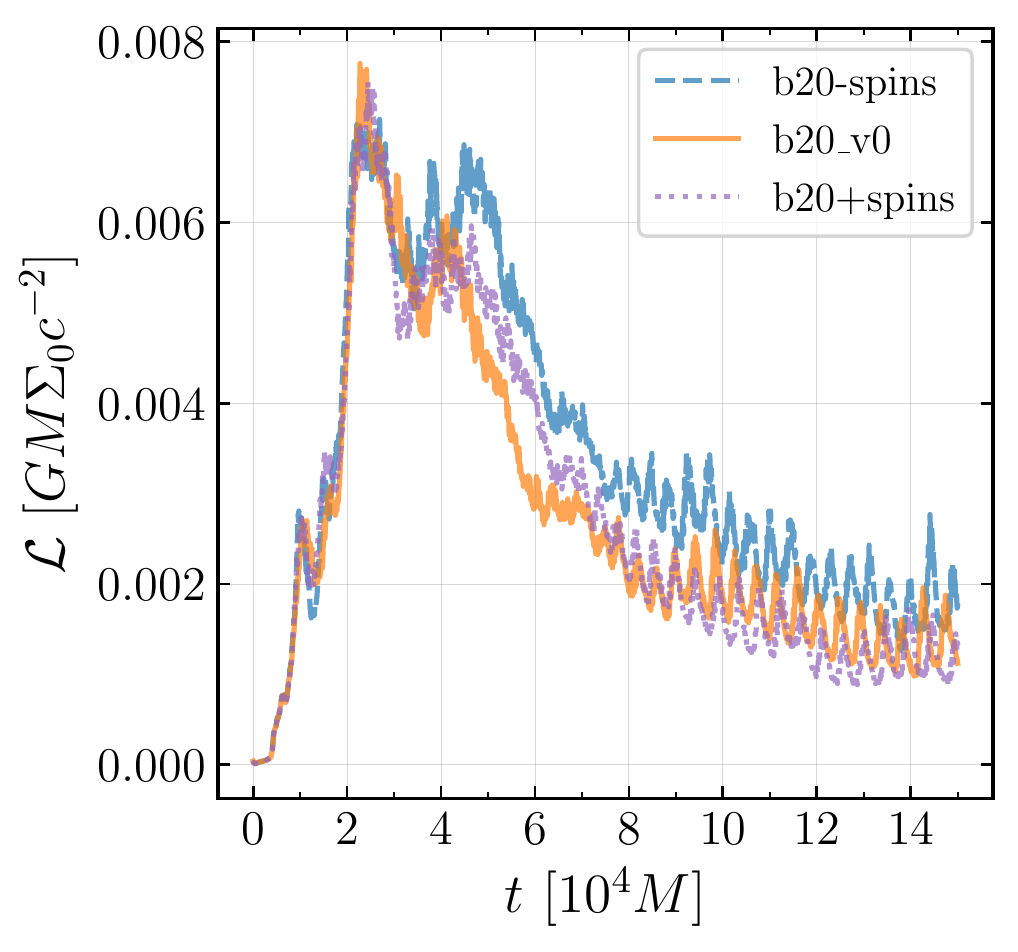}
  \caption{Integrated luminosity $\mathcal{L}$ (see Eq.~\eqref{e:luminosity}) as a function of time for three of our runs.}
  \label{f:luminosity}
\end{figure}

As found in previous works with similar parameters \citep{Shi2012, Noble2012}, a portion of the falling streams receives enough angular momentum from the binary and is flung back to the circumbinary disk, impacting the inner edge and causing strong shocks whose dissipation contributes significantly to the luminosity.
Having found that the accretion rate is sensitive to spin, we might therefore expect that the luminosity is likewise.
In particular, compared with non-spinning runs, the stronger streams of \runmspins{} should increase the total luminosity of the system, and the opposite for the weaker streams of \runpspins.
In Fig.~\ref{f:luminosity} we plot $\mathcal{L}$ as a function of time for our runs.
The average of $\mathcal{L}$ during the steady state period of non-spinning runs was $(1.76 \pm 0.07)\e{-3} G M \Sigma_0 c^{-2}$.
The corresponding averages for \runmspins{} and \runpspins{} depart  from this mean by $+7.49$ and $-3.17$ standard deviations, respectively, a very significant effect.
 These differences correspond to a change of $+29\%$ and $-12\%$ in the total luminosity of the system, respectively, with respect to non-spinning runs.

\begin{figure*}\centering
        \includegraphics[width=\linewidth]{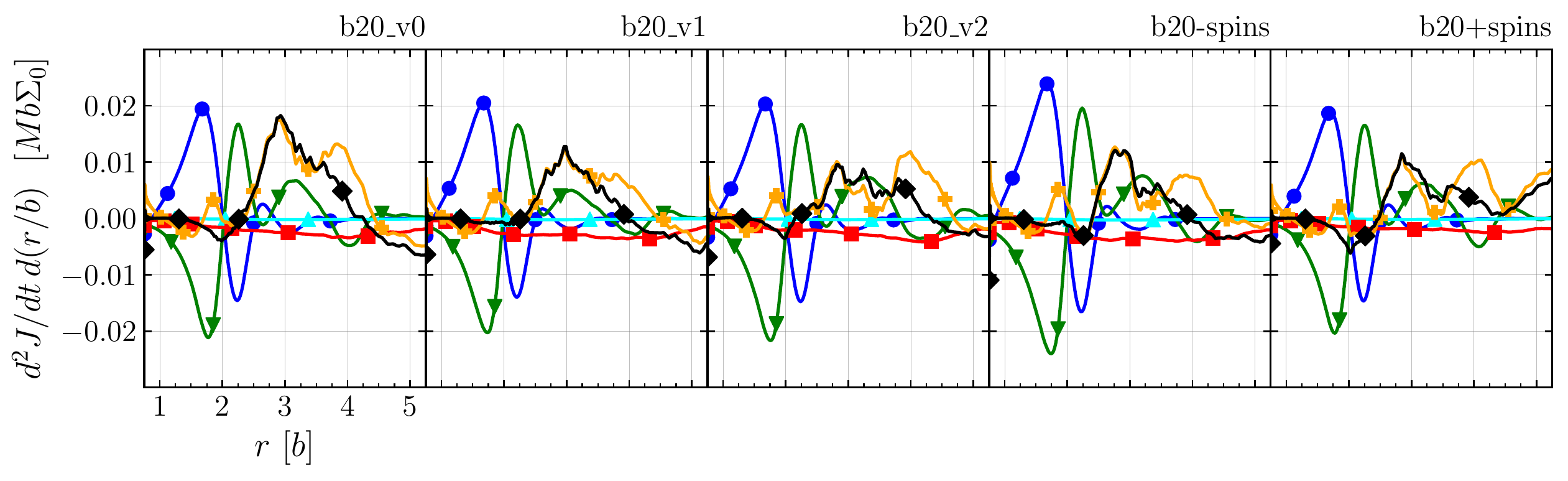}
	\caption{ Shell-integrated torques as a function of $r$, averaged over the period $t=70$ -- $150 \e{3} \, M$. We distinguish the gravitational torques excerted by the binary (\textit{blue/circle}), Maxwell stresses (\textit{red/square}), Reynolds or turbulent torques (\textit{green/down triangle}), the density of advected angular momentum (\textit{gold/plus}), radiative losses (\textit{cyan/triangle up}), and the net flux of angular momentum (\textit{black/diamond}).}
        \label{f:torques}
\end{figure*}

Besides carrying the accretion flow and driving shocks that contribute to the integrated luminosity, the streams also play an important role in angular momentum transport.   As  they plunge toward the binary, the streams are subjected to strong torques by the binary.  The portion of the stream flung back outward then transfers this added angular momentum to the inner edge of the circumbinary disk.
As explained by \cite{Shi2012}, because the local angular momentum $J=\int j^{t} \sqrt{-g} dV$ with $j^{\mu} = {T^{\mu}}_{\phi}$ should be constant in a time-steady disk, this supplemental angular momentum is transferred to adjacent layers by internal stresses.

To study the angular momentum budget of the circumbinary disk, we unpack $\partial_r \partial_t J$
into its several components.
We refer the reader to Appendix C of \cite{Noble2012} for the explicit expansion \citep[see also][]{Farris2011}.
Five stresses contribute:
the gravitational stress $T_{\mathrm{G}}$, whose radial gradient produces the gravitational torque ${T^{\mu}}_{\nu} {\Gamma^{\nu}}_{\mu \phi}$;
the Maxwell stress $\partial_r {M^{r}}_{\phi}$, which is the EM part of ${T^{r}}_{\phi}$;
turbulent Reynold stresses $\partial_r {R^{r}}_{\phi} = \rho \delta u^r \delta u^\phi$, resulting from local perturbations of the fluid velocity;
the advected Reynolds  stress $A^{r}_{\phi}$ associated with the mean velocities $u^r$ and $u^{\phi}$;
and the radiative stress $\mathcal{F}_{\phi}$ from the radiative cooling function.
Summed, these produce the local torque
\begin{eqnarray}
  \partial_r \partial_t J = \partial_r T_{\mathrm{G}} -  \left\{ \mathcal{F}_{\phi} \right\} - \partial_r \left\{ {M^r}_{\phi} \right\}
  \\ \nonumber
  - \partial_r \left\{ {R^{r}}_{\phi} \right\} - \partial_r \left\{ {A^r}_{\phi} \right\} \,,
  \label{e:dJdrdt}
\end{eqnarray}

In Fig.~\ref{f:torques} we plot each term on the RHS of Eq.~\eqref{e:dJdrdt} as a function of $r$, averaged over the period $t=70$ -- $150\e{3} M$.
The total angular momentum flux (\textit{black}) is approximately constant as a function of radius, as expected for a steady state flow.
In the cavity (i.e., $r < 2b$), there is a significant difference between our non-spinning and spinning runs.
The maximum of the gravitational torque (\textit{blue}) for non-spinning runs is $(2.011 \pm 0.053) \e{-2} M b \Sigma_0$, while \runmspins{} and \runpspins{} differ by $7.11$ and $-2.40$ standard deviations, respectively.
In other words, the stronger (weaker) streams from \runmspins{} (\runpspins) increase (reduce) the maximum of the gravitational torque density by $18\%$ ($-6\%$).
The Reynold stresses (\textit{green}) are increased (reduced) accordingly because once additional angular momentum is deposited by gravitational torques, it must be carried away by fluid motions.

\begin{figure*}\centering
        \includegraphics[width=.7\linewidth]{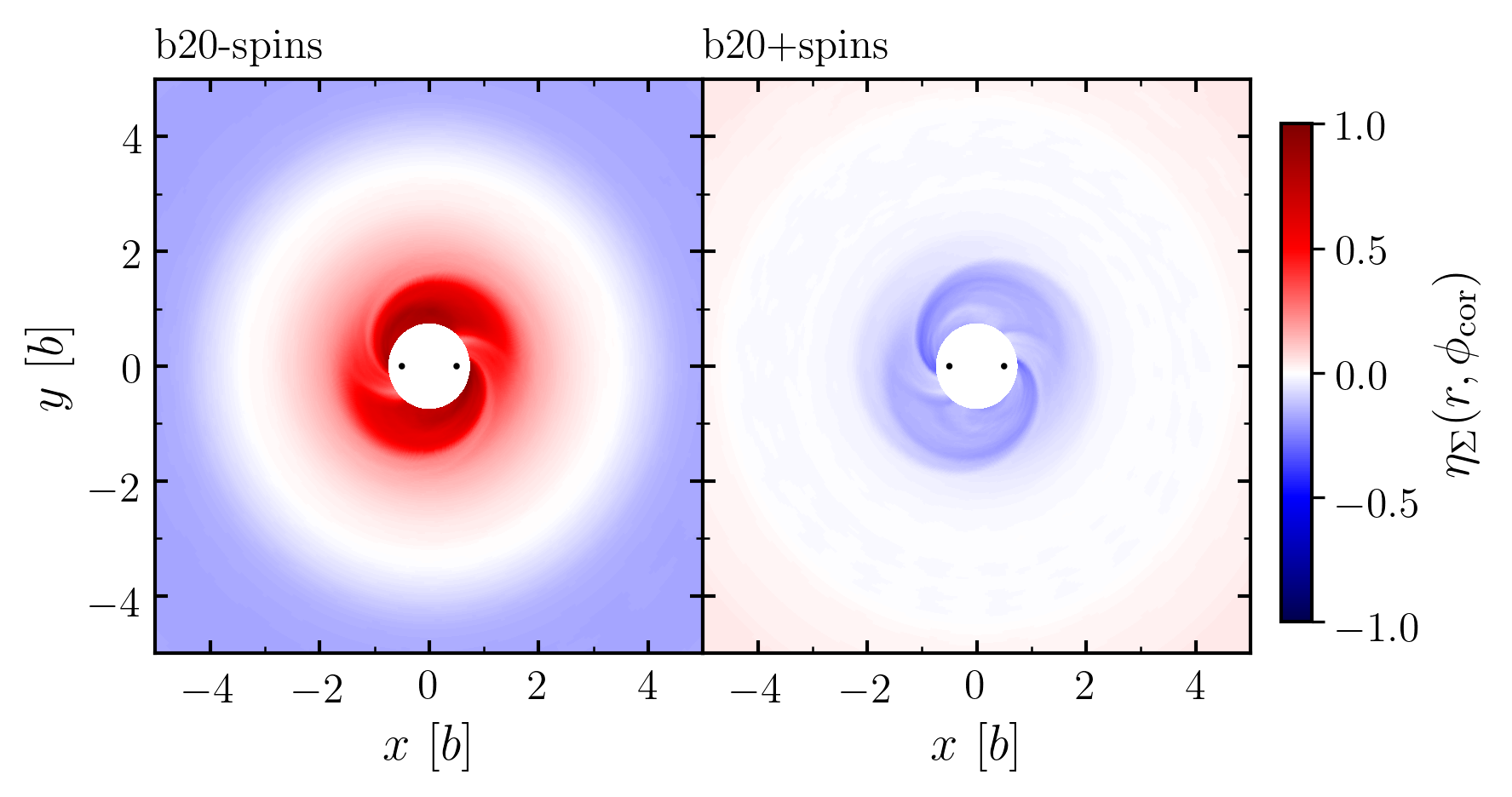}
        \caption{Residuals of the surface density with respect to non-spinning runs for  \runmspins{} (\textit{left}) and \runpspins{} (\textit{right}), averaged over the steady state period (see Eq.~\eqref{e:residuals}).}
        \label{f:Sigma_residuals}
\end{figure*}

\begin{figure}\centering
        \includegraphics[width=\linewidth]{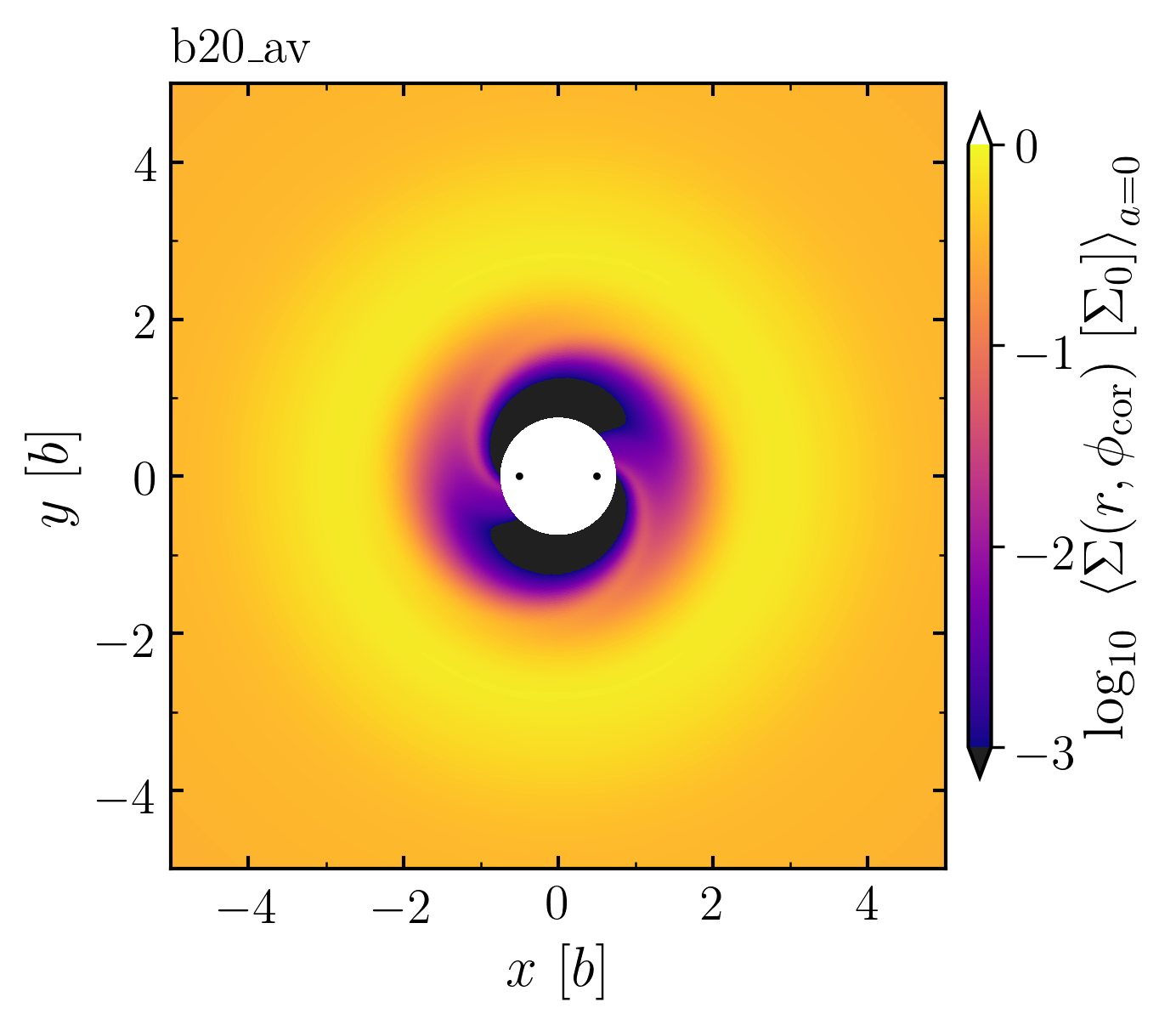}
        \caption{Average of the surface density over the steady state period for non-spinning runs, in the corotating frame of the binary (logarithmic scale).
        We notice the piling-up of matter in the inner region $2b<r<4b$, the evacuated inner cavity in $r<2b$, and the falling streams toward each BH.
        This function is the reference for the residuals in Fig.~\ref{f:Sigma_residuals}.}
        \label{f:Sigma_av}
\end{figure}

Contrasts in accretion rate must also, through mass conservation, affect the radial distribution of mass in the system.
To search for this effect, we contrast the surface density of gas $\Sigma(r,\,\phi)$ (see Eq.~\eqref{e:Sigma_rphi}) in the corotating frame of the binary with the surface density in non-spinning runs:

\begin{equation}
	\eta_{\Sigma}(r,\,\phi_{\mathrm{cor}}) = \frac{ \Sigma(r,\,\phi_{\mathrm{cor}}) [\Sigma_0] - \left< \Sigma(r,\,\phi_{\mathrm{cor}}) [\Sigma_0] \right>_{a=0} }{\left< \Sigma(r,\,\phi_{\mathrm{cor}}) [\Sigma_0] \right>_{a=0}} \, ,
  \label{e:residuals}
\end{equation}
where $\phi_{\mathrm{cor}}=\phi-\Omega_{\mathrm{bin}} t$, and the brackets $[\Sigma_0]$ denote that each surface density is taken in units of their initial maximum $\Sigma_0$.
In Fig.~\ref{f:Sigma_residuals} we plot the average of this residual over the steady state period for \runmspins{} (\textit{left}) and \runpspins{} (\textit{right}), and in Fig.~\ref{f:Sigma_av} we plot the averaged surface density in the corotating frame of the binary for non-spinning runs,
which is the reference function for the latter residuals.

The residual $\eta_{\Sigma}(r,\,\phi_{\mathrm{cor}})$ is greatest inside the cavity, where the spin effects should be largest.   The sign of the effect  is such that the residual in the cavity is positive for \runmspins{} but negative for  \runpspins{}---and flips in the bulk of the circumbinary disk ($r>4b$).
This is consistent with our results on enhanced (reduced) circumbinary accretion.
If, independent of spin, the system is in approximate
 inflow equilibrium, enhancement (or reduction) of the accretion rate implies that, averaged over time, the cavity must contain more (or less) gas mass for a fixed mass near the circumbinary disk's inner edge.  In addition, the outer disk is drained a bit more when there is a higher accretion rate at its inner edge when all the different initial disk masses were the same.

 \begin{figure}\centering
 	\includegraphics[width=.9\linewidth]{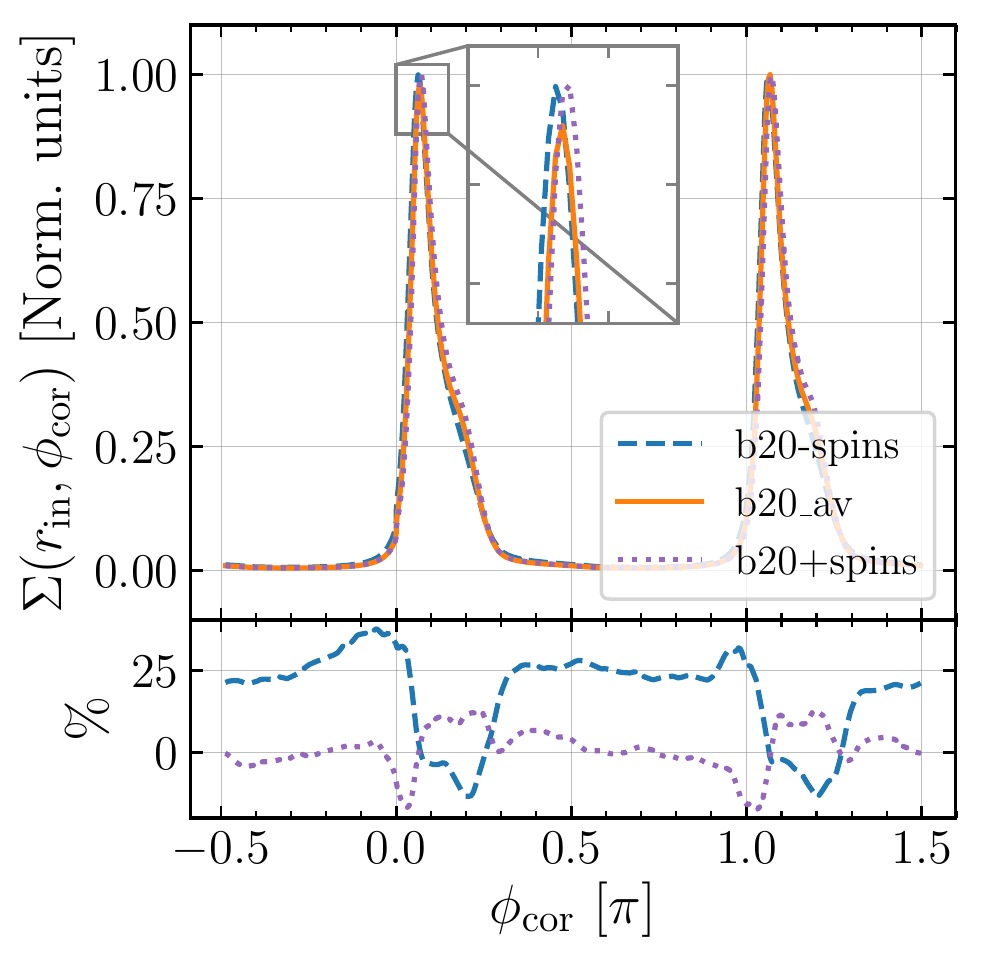}
         \caption{\textit{Top:} Surface density, averaged during the steady state period, in the corotating frame of the binary, and evaluated at the radial inner boundary of the domain $r_{\mathrm{in}}$, for spinning runs and the average of non-spinning runs (\texttt{b20\_av}). \textit{Bottom:} Persent deviation of curves for spinning runs with respect to the non-spinning average.}
         \label{f:SigmaAv_rin}
 \end{figure}

 Beyond the amount of matter and angular momentum that the streams carry, the spin of the BHs may affect the streams' trajectories.
 In Fig.~\ref{f:SigmaAv_rin} (\textit{top}) we plot the averaged surface density during the steady state period, in the corotating frame of the binary, evaluated at the radial inner boundary of the domain $r_{\mathrm{in}}$, for spinning runs and the average of non-spinning runs (\texttt{b20\_av}).
 We notice two distinctive peaks at $\phi_{\mathrm{cor}} \approx 0.06\pi, \, 1.06 \pi$ that we associate with the narrow streams that fall toward each BH (see also Fig~\ref{f:Sigma_av}).
 At a first glance, the curves for spinning and non-spinning runs look equivalent, but the zoom-in plot, and the percent deviations (\textit{bottom}), show some interesting results.
 First, we notice the peaks for the different runs are shifted in $\phi_{\mathrm{cor}}$, in ascending order \runmspins{}, \texttt{b20\_av}, \runpspins{}.
 Regarding the percent deviations of spinning runs, we notice \runmspins{} finds local maxima behind the peaks of the surface density, and local minima ahead, and \runpspins{} present the opposite behavior.
 In the next subsection, we explain these results from the effective gravitational potential of spinning binaries.

\subsection{Interpretation of Spin-Sensitive Results}
In Section~\ref{sec:spin-sens-results}  we found that the spin of the BHs in a binary system has significant effects on the circumbinary accretion and related quantities.
In this subsection, we explain the causes of these effects in simple physical terms.

In Appendix~\ref{s:postnewtonian}, we analyze the equations of motion of particles orbiting near spinning binaries.
We find the spin of the BHs introduces two effects to the lowest PN expansion of the gravitational potential.
First, the spin couples to the orbital velocity of the BH, as seen from the second and fourth terms of Eq.~\eqref{e:PNPhi}.
Far from the source and averaging in $\phi$, however, this effect is canceled for the case of identical BHs.
The second effect is frame-dragging, or twist of spacetime geodesics, as seen in Eq.~\eqref{e:PhiEff}.
Interestingly, this effect remains after expanding for large radius $r$ and averaging in $\phi$, and couples to the orbital angular momentum of the fluid.
In the following, we explore how this term affects the process of stream formation and accretion.

\cite{Shi2015} found that, in the phase-space of positions and velocities of the orbiting fluid, the volume of infalling trajectories from the inner edge of the circumbinary disk is severely constrained.
Gas with angular momentum close to the circular orbit angular momentum at the inner-edge radius falls in so slowly that the binary torques raise its angular momentum and the gas is cast back out to the circumbinary disk.   Only gas with angular momentum at least $\simeq 15\%$ less than that of a circular orbit can fall in quickly enough to avoid acquiring too much angular momentum.  Such gas parcels must, in addition, begin their fall from a specific angle relative to the binary separation axis.
The upper limit for the angular momentum $J$ of the fluid to be accreted is well approximated by the condition $\Phi_{\mathrm{eff}}(r_{\mathrm{in}}) \le 0$, where $\Phi_{\mathrm{eff}}$ is the gravitational effective potential of the binary, evaluated at the inner boundary of the domain.
As derived in Eq.~\eqref{e:PhiEff}:

\begin{equation}
  \Phi_{\mathrm{Eff}} =  -\frac{M}{r} - \frac{1}{16} \frac{b^2 M}{r^3} + \frac{J^2}{2r^2}
   + \frac{M J}{3r^{3}}\left(2 a + \frac{L}{4}\right) \, .
	\label{e:veff}
\end{equation}
The condition $\Phi_{\mathrm{eff}}(r_{\mathrm{in}}) \le 0$ is equivalent to $J \le (6.54, \, 6.51, \, 6.48)$ for $a=(-0.9, \, 0, \, 0.9)$, respectively.
In other words, spins opposite (parallel) to the angular momentum of the binary extend (reduce) the volume of infalling trajectories in the phase-space of position and velocity of the orbiting fluid.
This fact explains the enhanced (reduced) accretion in the run \runmspins{} (\runpspins).

In Fig.~\ref{f:SigmaAv_rin} we noticed the accretion streams for \runmspins{} (\runpspins) lie behind (ahead) in $\phi_{\mathrm{cor}}$ with respect to non-spinning runs.
In other words, the gas swings in azimuth by a smaller (larger) angle while traversing the cavity
before passing through the inner boundary.
This is also consistent with frame-dragging effects.

\subsection{Spin-Insensitive Results and Comparison with Previous Works}

In this subsection, we describe the properties of the circumbinary disk that are not significantly affected by spins, but the length of our simulations has revealed new aspects of them, not seen in previous, shorter simulations.

\begin{figure}\centering
  \includegraphics[width=.8\linewidth]{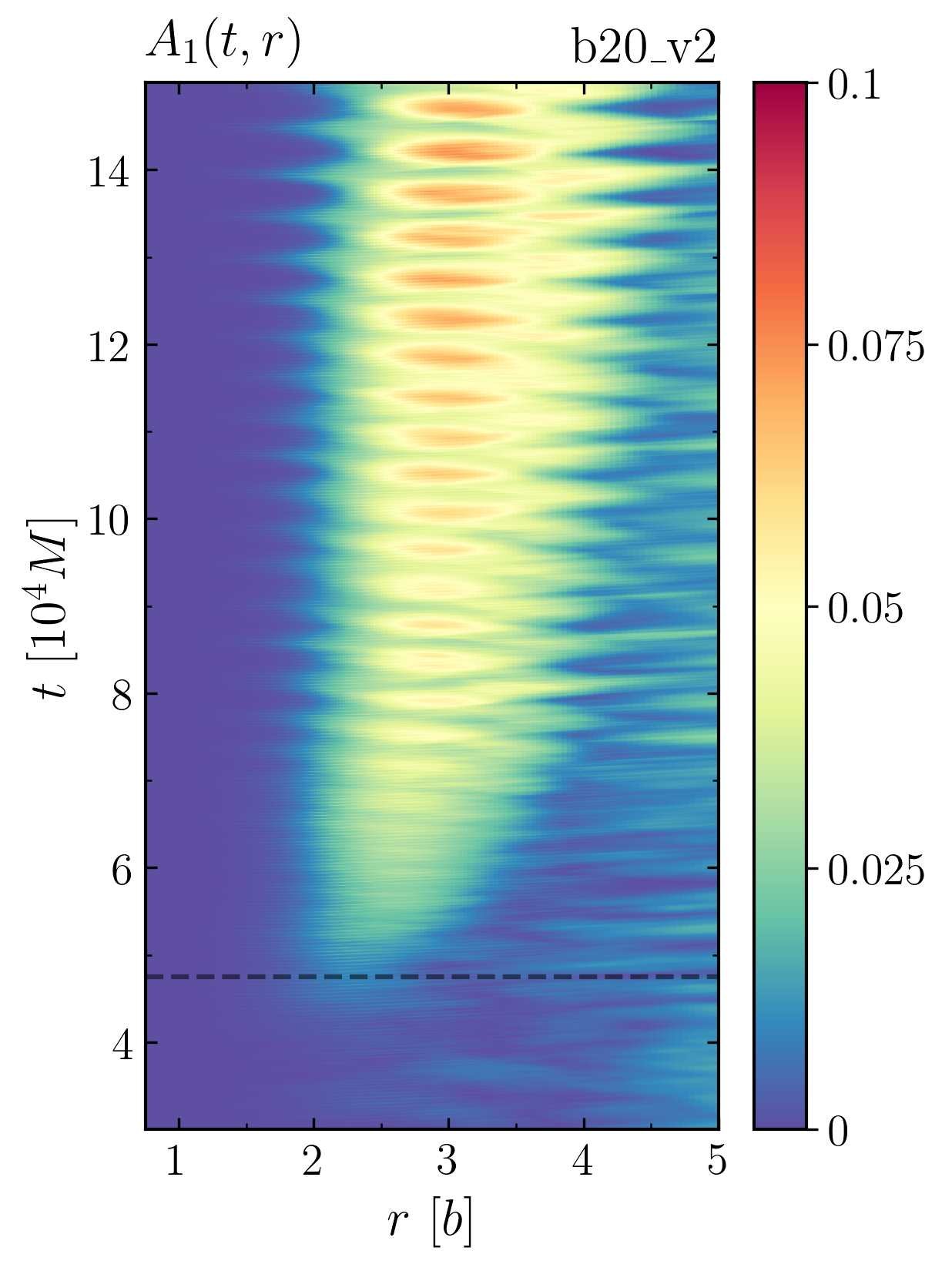}
  \caption{Power of $m=1$ mode of the vertically integrated density, as function of radii and time, for \runvt.
  We notice the growth and saturation of the lump at $2b<r<4b$.
  The \textit{dashed} line represents the moment of lump formation $t_{\mathrm{lump}}$.}
  \label{f:A1}
\end{figure}

    In binaries with mass-ratio close to unity and low orbital eccentricity, a remarkable $m = 1$ mode in the $\phi$-distribution of matter develops in the radial range $2b<r<4b$;  the so-called  \textit{lump}.  This lump arises as a result of phase-coherence in the trajectory of matter that falls a short way but then is propelled back out after the binary torques add to its angular momentum
\citep[see][]{Shi2012, Noble2012, DOrazio2013, Farris2014a, Miranda2017, Tang2017}.
As we will show, our longer simulations reveal that the dynamics of the lump are predictable from the time of its formation, and its orbit stabilizes after $\Delta t \sim 40\e{3}M$.

To  characterize the amplitude of the lump, we calculate the power of the Fourier modes $m=0$ and $m=1$ in the vertically integrated density as a function of radius and time (see Eq.~\eqref{e:mmode}, and \cite{Cuadra2009}, \cite{Noble2020}). We denote these modes $A_0(t,\,r)$ and $A_1(t,\,r)$, respectively.
In Fig.~\ref{f:A1}, we plot $A_1(t,\,r)$ for \runvt{} and, indeed, we notice the growth and saturation of the lump at $2b<r<4 b$.

\begin{figure}\centering
  \includegraphics[width=.8 \linewidth]{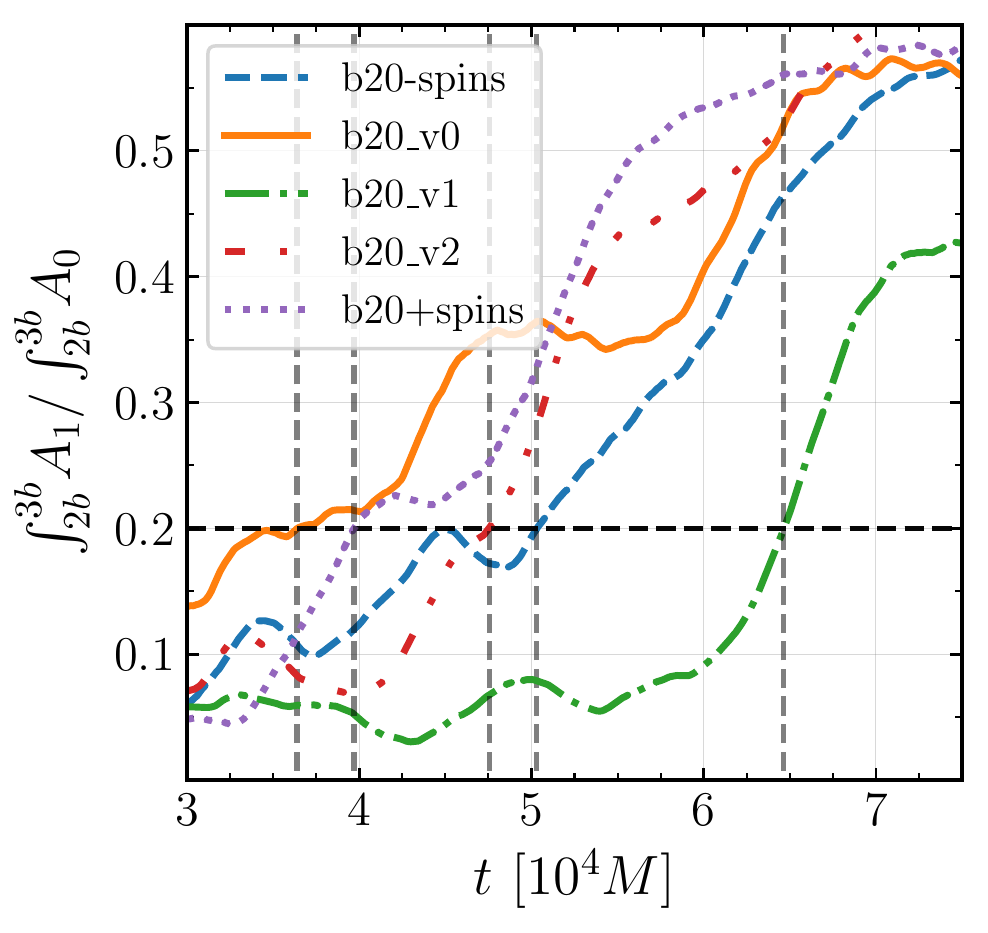}
  \caption{Evolution of the ratio of the power of the Fourier mode $m=1$ and $m=0$ of the vertically integrated surface at the lump formation region.
  The intersection of the black dashed and gray dashed lines defines the time of lump formation.
  In ascending order, $t_{\mathrm{lump}}/M= 36390, \, 39690, \, 47550, \, 50280, \,  64650$.}
  \label{f:lump_criteria}
\end{figure}

\begin{figure}\centering
        \includegraphics[width=\linewidth]{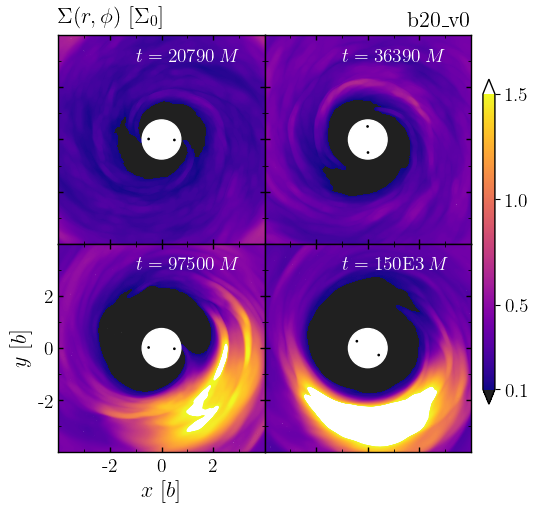}
        \caption{Surface density $\Sigma(r,\,\phi)$ in units of $\Sigma_0$ at different times for run \runvz.
        We distinguish the double accretion streams during the first orbits of the binary (\textit{top, left}), the formation of the lump (\textit{top, right}), the growth of the lump and the transition to a single accretion stream (\textit{bottom, left}), and the saturated lump at the end of the simulation (\textit{bottom, right}).}
        \label{f:Sigma_rphi_snapshots}
\end{figure}

To determine the time $t_{\mathrm{lump}}$ when the lump forms, we integrate $A_m(t,\, r)$ over the radial range $2b<r<3b$ and define $t_{\mathrm{lump}}$ as the time when the ratio of this integral of $A_1(t,\,r)$ to the total surface density (this integral of $A_0(t,\,r)$) is larger than $0.2$.
To visualize the different $t_{\mathrm{lump}}$ for each run, in Fig.~\ref{f:lump_criteria} we plot  the evolution of the ratio of the $m=1$ and $m=0$ integrals for our runs.
For non-spinning runs, the lump forms at $36390, \, 64650, \, 47550 \, M$, resulting in an average $t_{\mathrm{lump}} = (5.0 \pm 1.5) \e{4} M$.
In Fig.~\ref{f:Sigma_rphi_snapshots} (\textit{top, right}) we plot the surface density $\Sigma(r,\,\phi)$ (see Eq.~\eqref{e:Sigma_rphi}) at $t=t_{\mathrm{lump}}$ for \runvz{}, where we notice the recently formed lump in the positive $y$ hemisphere.
For runs \runmspins{} and \runpspins{}, the lump forms at $50280M$ and $39690M$ respectively, in concordance with non-spinning values.

\begin{figure}\centering
  \includegraphics[width=.8\linewidth]{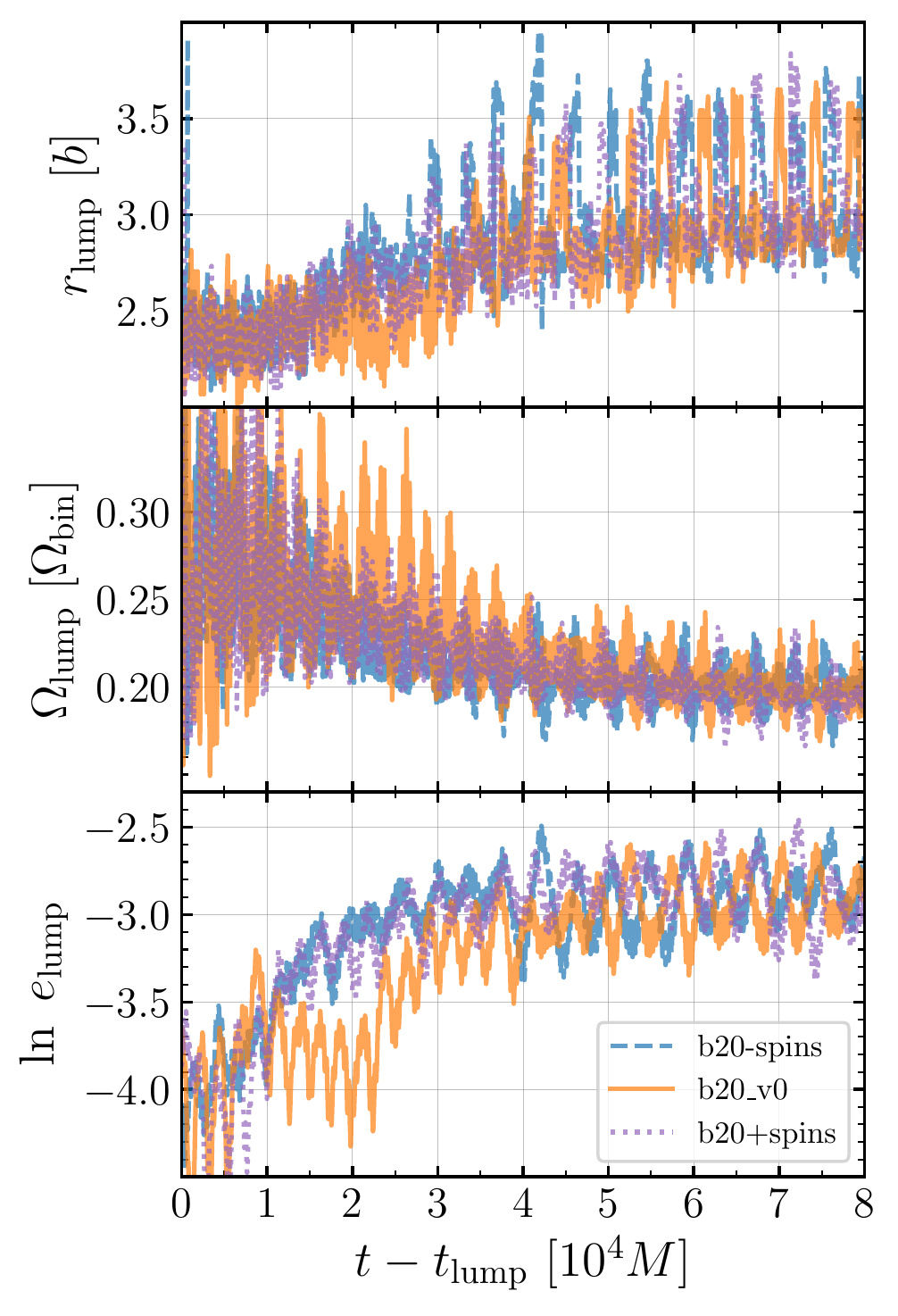}
 \caption{Evolution of the radial position of the lump (\textit{top}), its orbital frequency (\textit{middle}), and orbital eccentricity (\textit{bottom}).
 We plot the evolution of these quantities from the moment of lump formation $t_{\mathrm{lump}}$ until $t_{\mathrm{lump}}+8\e{4}M$.}
  \label{AllLump}
\end{figure}

To characterize the orbital dynamics of the lump,
we define $r_{\mathrm{lump}}(t)$ as the radial position of the maximum value of $A_1(t,\,r)$ as a function of time (see Fig.~\ref{f:A1}).   We define $\Omega_{\mathrm{lump}}(t)$
 in terms of the time-derivative of the Fourier phase for the $m=1$ mode (see Eq.~\eqref{e:omegalump}, and \cite{Noble2020}).
Lastly, we define the eccentricity $e_{\mathrm{lump}}(t)$ in terms of the ratio between the lump's radial and azimuthal velocity components (see Eq.~\eqref{e:eccentricity} and \cite{Shi2012}).
In Fig.~\ref{AllLump} we plot these quantities as a function of $t-t_{\mathrm{lump}}$, for each of our runs.
In each of the three panels, the curves follow very nearly the same paths.
This means that, although $t_{\mathrm{lump}}$ is subject to a considerable dispersion among our runs, once the lump forms, its dynamics are robust and predictable.

In Fig.~\ref{AllLump} we notice the orbit of the lump stabilizes after $\Delta t \sim 40\e{3}M$.
Its radial position $r_{\rm lump}$ grows approximately linear in time and then gradually equilibrates.
For the first $\Delta t = 20\e{3} M$, non-spinning runs present $ r_{\mathrm{lump}} = (2.48 \pm 0.05) \, b$, but for the last $\Delta t = 4 \e{3} M$ we find  $r_{\mathrm{lump}} = (3.05 \pm 0.05) \, b$.
The angular frequency of the lump decreases accordingly, following a Keplerian behavior.
For the first $\Delta t = 20\e{3} M$ we find $\Omega_{\mathrm{lump}} = (0.25 \pm 0.01) \, \Omega_{\mathrm{bin}}$ but for the last $\Delta t = 40 \e{3} M$ it reduces to $\Omega_{\mathrm{lump}} = (0.197 \pm 0.003) \, \Omega_{\mathrm{bin}}$.
This values are in agreement with previous works.
While the early value of $\Omega_{\mathrm{lump}}$ agrees with \cite{Noble2012} who evolved the system for the earlier stages of the lump development, the stabilized value of $\Omega_{\mathrm{lump}}$ agrees with longer two-dimensional hydrodynamical simulations (e.g., \cite{Miranda2017}).
Regarding the eccentricity, initially we find $\ln{e_{\mathrm{lump}}} = -3.6 \pm 0.2$ for non-spinning runs, but later it stabilizes to $\ln{e_{\mathrm{lump}}} = -2.91 \pm  0.04$, in agreement with results from \cite{Shi2012}.
Regarding our spinning runs, every quantity lies within $\pm 1.5$ times the standard deviation, implying the dynamics of the lump are independent of the spin of the BHs.

\begin{figure*}\centering
        \includegraphics[width=\linewidth]{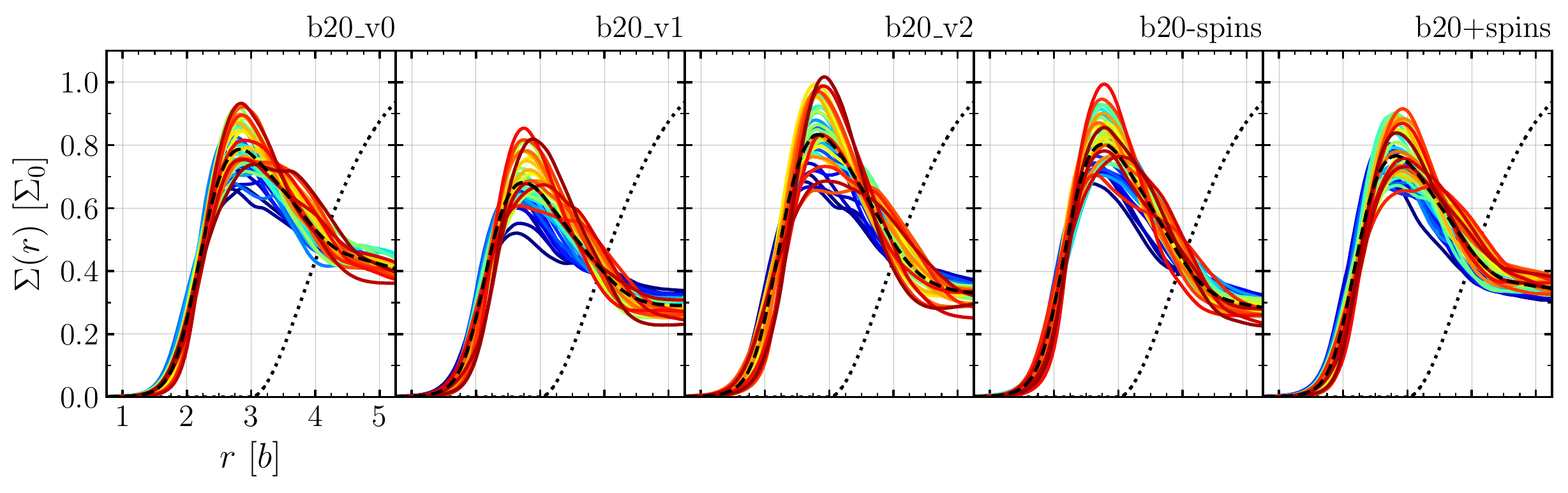}
        \caption{Vertically integrated and $\phi$-averaged density $\Sigma(r)$ averaged over $\Delta t = 2\e{3} M$ for the period $t=70$ -- $150\e{3} \, M$ (\textit{violet} to \textit{red} curves). The dotted curve represents the initial data, and the dashed curve the average of colored curves.}
        \label{f:Sigma_r}
\end{figure*}

In addition to the overdense lump, matter tends to pile up at the inner edge of the circumbinary disk.
This is as a consequence of the interplay between the internal stresses that remove angular momentum from matter just outside the inner edge and the gravitational torques that add angular momentum to streams in the cavity inside the inner edge.
 Our longer simulations reveal that this \textit{piling-up} saturates during the steady state.
To analyze the dynamics of this overdense region, in Fig.~\ref{f:Sigma_r} we plot the $\theta$-integrated and $\phi$-averaged density $\Sigma(r)$ (Eq. \eqref{e:Sigma_r}) in the period $t=70$ -- $150\e{3} M$.
Each curve represents an average over $\Delta t = 2\e{3}M$ and, through colors violet to red, they span the entire steady state period.
The maximum of the averaged curves (\textit{dashed}) for non-spinning runs is $(0.76 \pm 0.07) \, \Sigma_0$, and its radial position is $(2.80 \pm 0.06) \, b$.
\cite{Noble2012} found that, when the binary evolution is frozen, the peak surface density increased steadily up to $t \approx 75\e{3}M$; Fig.~\ref{f:Sigma_r} reveals that their simulation stopped just at the point where the growth in peak surface density ceases.

\begin{figure*}\centering
	\includegraphics[width=.35\linewidth]{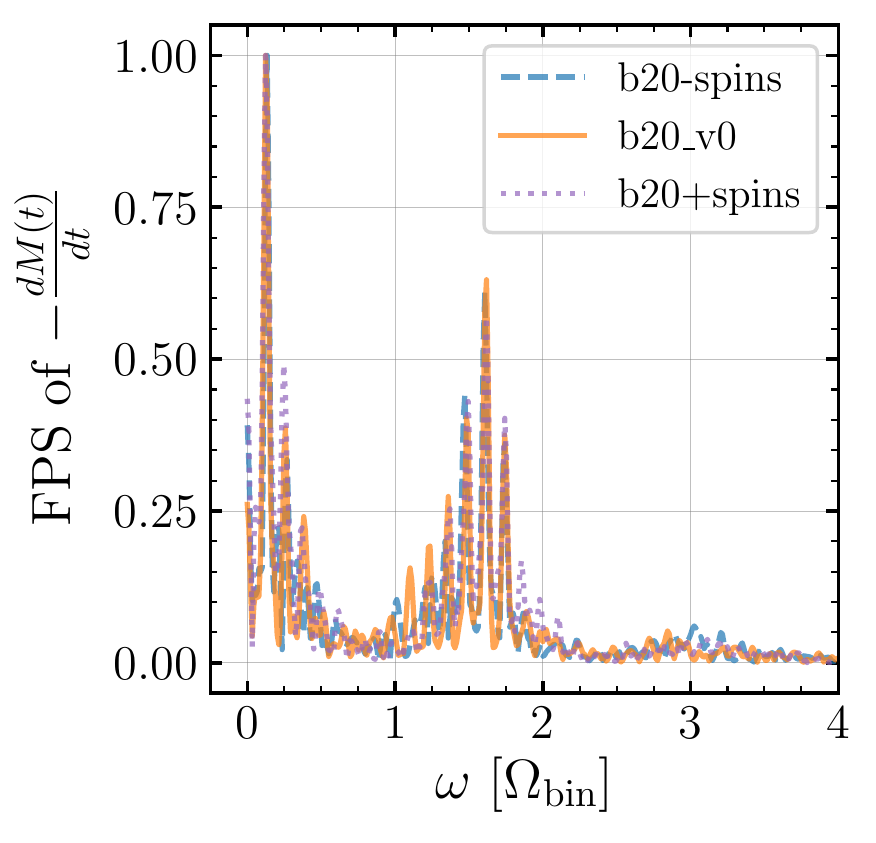}
  \includegraphics[width=.35\linewidth]{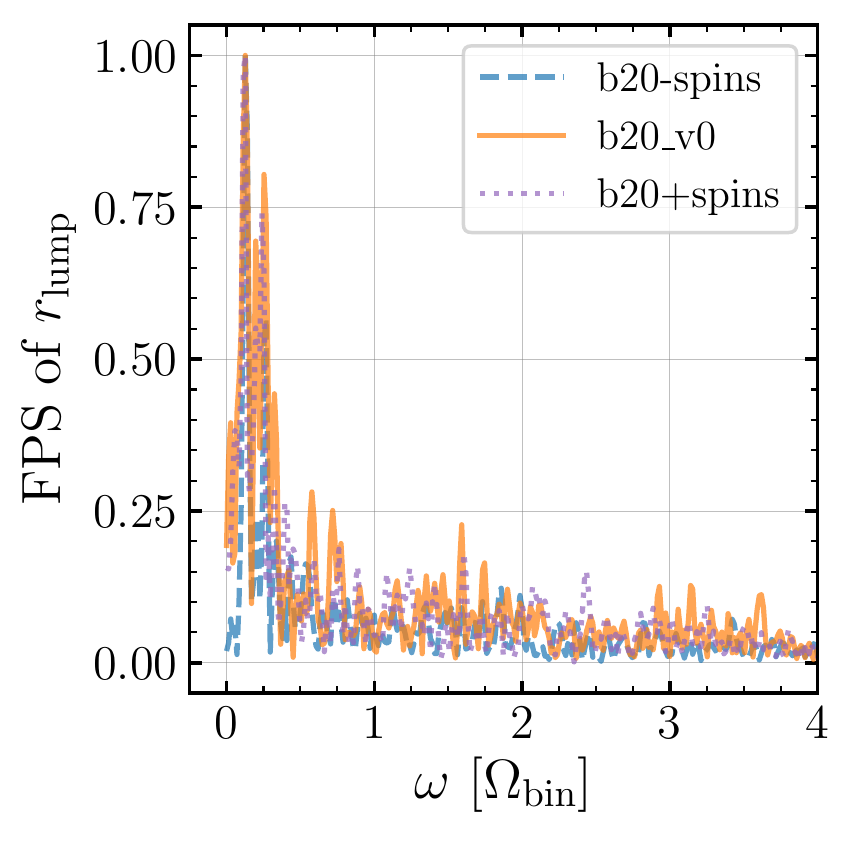}
  \caption{Fourier power spectrum of the accretion rate at the innermost radial boundary of the domain during the steady state period (\textit{left}), and of the radial position of the lump in the interval $40\e{3}\, M<t-t_{\mathrm{lump}}<80\e{3}\, M$ (\textit{right}).
  To enhance the periodic behavior, we analyze the difference of these quantities with adjusted polynomials of first order, and we apply a \textit{Blackman-Harris} window function.
  }
  \label{f:FPS}
\end{figure*}

The length of our simulations makes them ideal to study the characteristic frequencies of the system.
As expected from the predominant quadrupole mode of the binary potential, accretion into the inner boundary is initially carried by two narrow streams that extend from the inner edge of the disks toward each BH (see Fig.~\ref{f:Sigma_rphi_snapshots} (\textit{top, left})). In the steady state, however, accretion is dominated by a single stream that is produced when a BH passes near the lump (see Fig.~\ref{f:Sigma_rphi_snapshots} (\textit{bottom, left})).
The frequency of this occurrence is $2 (  \Omega_{\mathrm{bin}} - \Omega_{\mathrm{lump}} ) \sim 1.6  \Omega_{\mathrm{bin}}$.
Indeed, in Fig.~\ref{f:FPS} (\textit{left}) we plot the Fourier Power Spectrum (FPS) of the accretion rate during the steady state (see also Fig.~\ref{f:Mdot}) and find the expected peak at $\omega =  1.6  \Omega_{\mathrm{bin}}$.
Fig.~\ref{f:FPS} (\textit{left}) also presents a strong peak at $\omega=0.12 \Omega_{\mathrm{bin}}$, which corresponds to a lower-frequency modulation of the accretion rate, as seen in the spikes of Fig.~\ref{f:Mdot} during the steady state.

This modulation is caused by an oscillation of the radial position of the lump.
Indeed, in Fig.~\ref{f:FPS} (\textit{right}) we plot the FPS of $r_{\mathrm{lump}}(t)$ during the stabilized period of the orbit of the lump $40\e{3}\, M<t-t_{\mathrm{lump}}<80\e{3}\, M$ (see also Fig.~\eqref{AllLump} (\textit{top})), and find its maximum at $\omega=0.12 \Omega_{\mathrm{bin}}$.
The causes of this radial oscillation will be addressed in subsequent work.

Although the bimodal distribution of the FPS of the accretion rate at the cavity is in agreement with previous works \citep[][among others]{MacFadyen2008, Shi2012, DOrazio2013, Munoz2016}, the precise values of the peak frequencies claimed in this work differ with such references.

\begin{figure*}\centering
        \includegraphics[width=\linewidth]{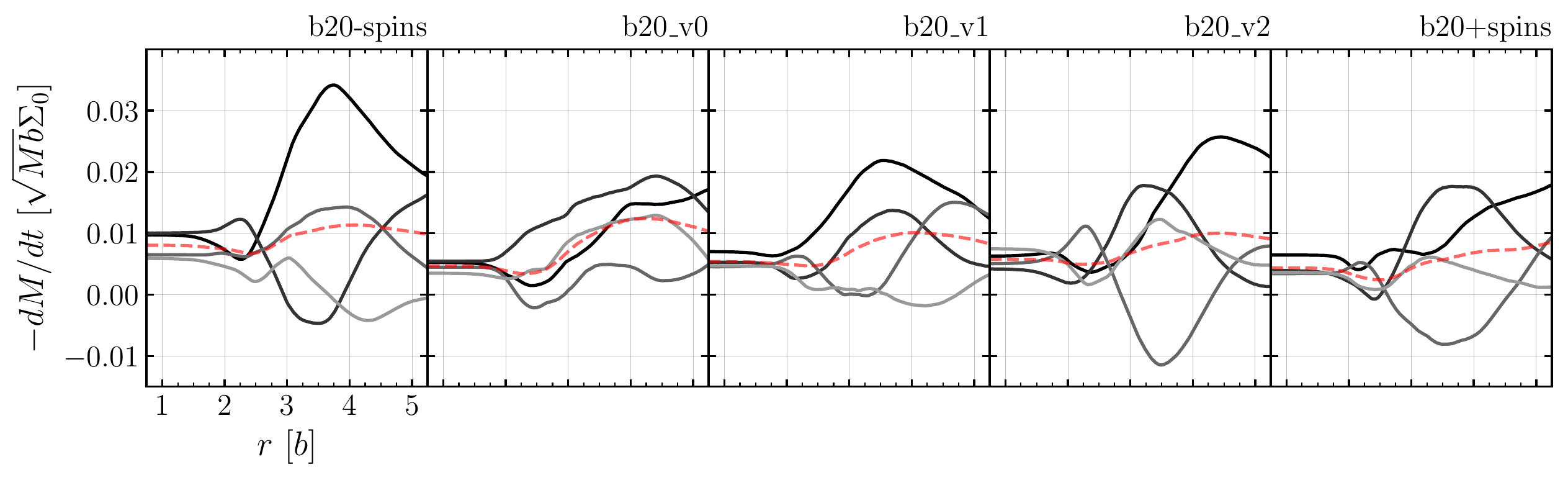}
        \caption{\textit{Top:} Accretion rate as a function of $r$ averaged over $\Delta t = 20\e{3}M$ in the period $t=70$ -- $150\e{3} M$ (\textit{dark} to \textit{light} curves), and the average over the full period (\textit{dashed, red}).}
        \label{f:mdot}
\end{figure*}

In the following, we analyze if our longer simulations approach inflow equilibrium.
In Fig.~\ref{f:mdot} we plot the average of the radial profile of the accretion rate (see Eq. \eqref{e:mdot}) over the period $t=70-150\e{3} M$ (\textit{dashed, red}), and over four equally spaced sub-periods with $\Delta t = 20 \e{3} M$ (\textit{dark} to \textit{light} curves).
While we notice the average curve has a systematic growth from $r>3b$, implying the disk has not reached inflow equilibrium, we also notice an improvement of this inflow condition if compared with previous three-dimensional MHD simulations that evolved the system for earlier stages \citep{Shi2012, Noble2012}.
In those earlier papers, the ratio of the accretion rate at $r \simeq 3 - 4b$ to the accretion rate crossing the inner edge was about a factor of 3; extending the duration of the simulation from $70\e{3}M$ to $150\e{3}M$ reduces that contrast to a factor $\approx 1.5$.  Thus, for these conditions, the inflow equilibration time at this radius is $t\sim 10^5M$.

Interestingly, also in Fig.~\ref{f:mdot}, we notice periods when the accretion rate becomes negative at regions close to the inner edge of the disk.  In other words, the inner edge presents periods  of slowly outward receding.

Finally, we analyze the mechanisms of angular momentum transport during the steady state of our longer runs. Comparing our results on shell-integrated torques in Fig.~\ref{f:torques} with those of \cite{Shi2012} and \cite{Noble2012} for earlier epochs of the system, we find that during the steady stage Maxwell stresses (\textit{red}) are significantly reduced at the bulk of the disk.
On the contrary, Reynolds stresses are strengthened and show traces of a propagating wave through the fluid.  \cite{Shi2015} also found a significant growth of turbulent torques in the steady state period of a related system, and associated it with the propagation of a single-armed wave.
We will study these issues in detail in a subsequent work.

\subsection{Summary}

We found the spin of the BHs have a direct impact on processes that take place in the inner cavity.
Specifically, negative (positive) spins enhance (reduce) the circumbinary accretion by $+45\%$ ($-14\%$) with respect to the non-spinning case.
The stronger (weaker) accretion streams enhance (decrease) the globally integrated luminosity by $+29\%$ ($-12\%$), and the peak of the gravitational torques at the cavity by $+18\%$ ($-6\%$), with respect to non-spinning runs.
In the long-term, these effects can discreetly influence the bulk of the disk, as we found  a reduction (increment) of the surface density for $r>4b$.
Finally, we found that the spin of the BH affects the shape of the accretion streams, as they fall behind (advanced) in $\phi$ with respect to the non-spinning case.
Other properties of the circumbinary such as the dynamics of the lump, the piling-up of matter at the inner edge, or the radial profile of the accretion rate, remain unaffected by the spin of the BHs.

The length of our runs allowed us to reach an unprecedented steady state in three-dimensional GRMHD simulations.
We found interesting differences with respect to early epochs, as described by \cite{Shi2012} and \cite{Noble2012}.
In particular, the dynamics of the lump are robust and predictable since its formation and its orbit stabilizes after $\Delta t \sim 40\e{3}M$, the growth of the lump and the piling up of matter at the inner edge of the disk saturates and remains steady, there are regions within the inner part of the circumbinary where the accretion rate becomes negative, and the role of Reynolds stresses grow at the bulk of the disk, but the role of Maxwell stresses diminishes, revealing a demagnetization of the plasma.
We intend to explore these results in detail in an upcoming work.

\section{Conclusions}
\label{sec:conclusions}

We presented a new approximate metric for the spacetime of widely separated BBHs in the relativistic regime ($b\ge 20M$).
This metric is unique in the sense that it can be used in the strong field regime, is easy to implement, it has an optimal performance, and includes the spins of the BHs as free parameters.
We computed and analyzed its Ricci scalar $R$ at different scales and concluded the metric is an acceptable approximation for a vacuum solution of EFE.
Further, we proved its expansion agrees with the lowest PN expansion of the metric of a binary system of spinning BHs.

As a first application, we set and evolved a series of magnetized circumbinary disks around an equal-mass binary system, with separation fixed at $b=20M$.
We explored different values for the spins of the BHs, aligned and counter-aligned with the orbital angular momentum of the binary, and performed three identical non-spinning runs to study the effect of random perturbations in our predictions (see Table~\ref{t:runs}).
We followed closely the techniques of \cite{Noble2012} that explored the same system (non-spinning) but for earlier stages and with a different approach and gauge for the spacetime construction.
We evolved the system for longer than previous three-dimensional MHD simulations, until $t=150\e{3}M$ or $266$ orbits of the binary system.
We noticed that the circumbinary disk reaches a steady state from $t=75\e{3}M$ onwards, and focused our results on this period.
Our results are consistent with previous works on non-spinning binaries, and with expectations for the effect of the spins on the circumbinary disks, proving the physical validity of the SKS spacetime.
We conclude the spin of the BHs, via frame-dragging effects, can significantly affect the circumbinary accretion and luminosity.
Specifically, spins counter-aligned (aligned) with the orbital angular momentum of the binary enhance (reduce) the circumbinary accretion with respect to the non-spinning case.
Further, the spin twists the spacetime geodesics at the cavity, and the streams reach the inner cavity behind (forward) in $\phi$ with respect to the non-spinning case.
We will explore the dynamics of mini-disks around spinning binaries in a subsequent work \citep{Combi2021}.

\acknowledgments

F.~L.~A., M.~C. and M.~A. acknowledge support from the National
Science Foundation (NSF) from Grants Nos AST-2009330, AST-1028087, AST-1516150 and PHY-1707946. Partial support
for F.~L.~A. and M.~C. was also provided by NASA TCAN grant No. 80NSSC18K1488, PHY-1607520, and PHY-1305730.
F.~L.~A. also acknowledges support from a CONICET fellowship.
L.~C. acknowledges support from the RIT's Frontier of Gravitational Wave Astronomy center, and by the CONICET fellowship.
M.~A. received support from RIT's Frontiers in Gravitational Wave Astrophysics Postdoctoral Fellowship.

S.~C.~N. was supported by AST-1028087, AST-1515982 and
OAC-1515969, and by an appointment to the NASA Postdoctoral Program at
the Goddard Space Flight Center administrated by USRA through a
contract with NASA.

D.~B.~B. received support from NSF Grants Nos AST-1028087, AST-1516150, PHY-1707946, and
by the US Department of Energy through the Los Alamos
National Laboratory. Los Alamos National Laboratory is operated by Triad National
Security, LLC, for the National Nuclear Security Administration of U.S.
Department of Energy (Contract No. 89233218CNA000001).

V.~M. received support from NSF Grants Nos OAC-1550436, AST-1516150, and by the Exascale Computing
Project (17-SC-20-SC), a collaborative effort of the U.S. Department of Energy
(DOE) Office of Science and the National Nuclear Security Administration. Work at
Oak Ridge National Laboratory is supported under contract DE-AC05-00OR22725 with the
U.S. Department of Energy.

H.~N. acknowledges support from JSPS KAKENHI Grant Nos. JP16K05347 and JP17H06358.

J.~H.~K. was partially supported by NSF Grant PHYS-1707826 and AST-AST-2009260.

Computational resources were provided by the NCSA's Blue Waters
sustained-petascale computing NSF projects OAC-1811228 and OAC-1516125, and by the TACC's Frontera NSF projects  PHY20010 and AST20021. Additional local resources were provided by the RIT's BlueSky and Green Pairie Clusters
acquired with NSF grants AST-1028087, PHY-0722703, PHY-1229173 and PHY-1726215.

%





\newpage

\appendix

\section{Diagnostics}
\label{sec:Diagnostics}
We summarize usual conventions and quantities used in the analysis of accretion disks:

\begin{itemize}

  \item Surface density:
  
  \begin{equation}
    \label{e:Sigma_rphi}
    \Sigma(r,\phi) =\frac{\int \rho \sqrt{-g} \, d\theta}{\sqrt{g_{\phi \phi}(\theta=\pi / 2)}} \,,
  \end{equation}
  and its $\phi$-average:
  
  \begin{equation}
    \label{e:Sigma_r}
    \Sigma(r)
    = \frac{\int \rho \sqrt{-g} \, d\theta \, d\phi}{\int \sqrt{g_{\phi \phi}(\theta=\pi / 2)} \, d\phi} \,.
  \end{equation}

  \item Height of the disk:
  
    \begin{equation}
      \label{e:height}
      H = \frac{\left< \rho \sqrt{g_{\theta \theta}} \left| \theta  - \pi /2 \right| \right>}{\left< \rho \right>} \,.
    \end{equation}

  \item Accretion rate as a function of $r$:
  
  \begin{equation}
    \label{e:mdot}
    \frac{dM(r)}{dt} = - \int \rho u^r \sqrt{-g} \, d\theta \, d\phi \,.
  \end{equation}

  \item Eccentricity of the fluid in the lump region:
  
    \begin{equation}
      \label{e:eccentricity}
      e_{\mathrm{lump}} = \frac{\left| \int_{2b}^{4b}\, \left< h v_r e^{i \phi} \right> \, dr \right|}{\int_{2b}^{4b}  \left< h v_{\phi} \right>\, dr} \,.
    \end{equation}

  \item Integrated luminosity:
  
  \begin{equation}
    \label{e:luminosity}
    \mathcal{L} = \int u_t \mathcal{L}_c \sqrt{-g} \, dr \, d\theta \, d\phi \,.
  \end{equation}

  \item Fourier transformation with respect to $\phi$ of the vertically integrated density:
  
  \begin{equation}
    \mathcal{B}_m(r,\,t) = \left\{\rho e^{i m\phi}\right\}\, ,
  \end{equation}
  and the corresponding mode power
  
  \begin{equation}
    \label{e:mmode}
    A_m(r,t) = \left|   \mathcal{B}_m(r,t) \right| \,.
  \end{equation}

  \item Since the vertically integrated surface is a real function, its Fourier modes satisfy:
  
  \begin{equation}
     \mathcal{B}_m (r,\,t) e^{i m \phi}
     =  \mathrm{Re}\left[\mathcal{B}_m(r,\,t)\right] \cos(\phi) - \mathrm{Im}\left[\mathcal{B}_m(r,\,t)\right] \sin(\phi) \,.
  \end{equation}
  The maximum of these modes is found at the phase:
  
  \begin{equation}
    \phi_m(r,\, t) = \arctan\left[ -\frac{\mathrm{Im}\left[\mathcal{B}_m(r,\,t)\right]}{\mathrm{Re}\left[\mathcal{B}_m(r,\,t)\right]}\right]\, .
  \end{equation}
  where the function $\arctan$ is defined to return the value of $\phi_m$ in the range $[0,2\pi]$, checking on the sings of $\mathrm{Im}\left[\mathcal{B}_m(r,\,t)\right]$ and
  $\mathrm{Re}\left[\mathcal{B}_m(r,\,t)\right]$.
  Identifying the phase of the lump as the maximum of the $m=1$ mode integrated at the lump region $2b<r<4b$, we obtain the angular frequency of the lump:
  
  \begin{equation}
    \label{e:omegalump}
    \Omega_{\mathrm{lump}} = \frac{d}{dt} \phi_{m=1} (t)\,.
  \end{equation}

\end{itemize}

\section{Post-Newtonian approximation}
\label{s:postnewtonian}
  The linearity of the SKS metric \eqref{eq:SKSmetric} implies that its expansion to the lowest order in the boost velocities and large radius, reproduces the lowest PN metric of a spinning compact binaries \citep[see][]{Tagoshi2001}:
  \begin{eqnarray}
  		g_{00} &=& -(1 +2\Phi) + \mathcal{O}(v^2_{\mathrm{K}}) \,,
\\
  		g_{ij} &=& \delta_{ij}(1 - 2\Phi)+ \mathcal{O}(v^2_{\mathrm{K}}) \,,
\\
      g_{0i} &=& -4\xi_i+ \mathcal{O}(v^2_{\mathrm{K}}) \,,
  \end{eqnarray}
  where the scalar and vector potentials read:
  \begin{eqnarray}
  		\Phi &:=& -\frac{M^{(1)}}{r^{(1)}}- 2\frac{S^{(1)ij} \check{r}^{(1)i} v_{\mathrm{K}}^{(1)j}}{r^{(1)2}} + \left[(1)\rightarrow (2)\right],
\\
  		\xi_i &:=& \frac{M^{(1)}}{r^{(1)}} v^{(1)i}+ 2\frac{\check{r}^{(1)k}}{r^{(1)2}} S^{(1)ki} + \left[(1)\rightarrow (2)\right],
  \end{eqnarray}
  being $S^{(n)ij}:= \epsilon^{ijk}S^{(n)k}$ with $S^{(n)k}:= \delta^k_z M^{(n)} a^{(n)}$ for spins aligned with the $z$-axis, $\vec{r}^{(n)} = (x^{(n)},y^{(n)},z^{(n)})$ the positional vector referred to the $n$-th BH frame, $\check{r}^{(n)i}$=$r^{(n)i}$/$r^{(n)}$, and $v_{\mathrm{K}}^{(n)i}$ the spatial velocity of the $n$-th BH.

In the PN approximation, the equations of motion for a test particle with velocity $\vec{V}$ are given by the so called gravitomagnetic analogue of Lorentz equations \citep[see, for instance,][]{Mashoon2003}:

  \begin{equation}
    \label{eq:eqmotion}
  		\frac{dV^i}{dt} = - \nabla^i \Phi - (\vec{V}\times(\nabla\times \vec{\xi}))^i \,.
  \end{equation}
In the following we derive the dominant terms in these equations for the case of identical BHs orbiting in circular orbits, and we discuss the effect of the spins.

The scalar potential can be written as:
  \begin{eqnarray}
  		\Phi =-\frac{M^{(1)}}{r^{(1)}} + 2 \frac{\check{r}^{(1)}\cdot (\vec{S}^{(1)}\times \vec{v}_{\mathrm{K}}^{(1)})}{r^{(1)2}}  + \left[(1) \rightarrow (2) \right] \\
  				= -\frac{M^{(1)}}{r^{(1)}} + 2\frac{M^{(1)} a^{(1)} v_{\mathrm{K}}^{(1)} \cos\phi^{(1)} }{r^{(1)2}}
 + \left[(1) \rightarrow (2) \right] \,,
    \label{e:PNPhi}
\end{eqnarray}
where $\vec{S}^{(n)} \times \vec{v}_{\mathrm{K}}^{(n)}$ points towards the center of mass and
thus $\phi^{(n)}$ is the angle between $\check{r}^{(n)}$ and $-\check{r}^{(n)}_{\mathrm{K}}$,
being $\check{r}^{(n)}_{\mathrm{K}}=\vec{r}_{\mathrm{K}}^{(n)}/|\vec{r}_{\mathrm{K}}^{(n)}|$ the normalized vector of the position of the $n$-th BH
(see Eqs.~\eqref{traj1} and \eqref{traj2}).
The second and fourth terms in Eq.~\eqref{e:PNPhi} represent spin-orbit coupling effects.

Taking into account the following expressions:
\begin{eqnarray}
  {r^{(1,2)}}^2 &=& r^2 + \frac{1}{4} b^2 \mp r b \cos\left(\phi - \Omega_{\mathrm{bin}} t\right)\,, \\
  \cos \phi^{(1,2)} &=& \frac{\pm x_{\mathrm{K}}^{(1,2)} \mp x}{r^{(1,2)}},
\end{eqnarray}
where $b$ is the separation of the binary, $\vec{r}=(x,y,z)$ is the position vector of the particle in Cartesian coordinates with respect to the center of mass of the binary, and $\phi=\arctan(y/x)$, we expand Eq.~\eqref{e:PNPhi} for large radius $r$ and average in $\phi$ and $t$, to obtain:

\begin{equation}
    \Phi =-\frac{M}{r} - \frac{1}{16} \frac{b^2 M}{r^3} + \mathcal{O}\left(\frac{1}{r^4}\right)\, ,
  \label{e:PNPhiInf}
\end{equation}
where $M=M^{(1)} + M^{(2)}$.
In this equation we recognize the well-known multipole expansion of equal-mass binaries in the Newtonian regime \cite[see, for instance, ][]{MacFadyen2008}.
We also notice that spin-orbit coupling terms canceled out in this limit, because of the symmetries of the system.

Regarding the vector potential $\xi_i$, we obtain the Cartesian components:

\begin{equation}
  \vec{\xi} = \frac{M^{(1)}}{r^{(1)}} \vec{v}^{(1)} + 2 \frac{a^{(1)} M^{(1)}}{{r^{(1)}}^{3}}\left(y^{(1)} ,\,-x^{(1)} , 0\right)
  + \left[(1) \rightarrow (2) \right] \, .
\end{equation}
Replacing this expression in the second term of Eq.~\eqref{eq:eqmotion}, transforming to a spherical basis, expanding for large radius $r$, and averaging in $\phi$, we obtain the radial component:

\begin{equation}
  (\vec{V}\times(\nabla\times \vec{\xi}))^r =  \frac{M J}{r^{4}}\left(2 a + \frac{L}{4}\right)
   + \mathcal{O}\left(\frac{1}{r^5}\right)\,,
  \label{e:PNXiInf}
\end{equation}
where we have defined $J= x V^y - y V^x$  as the specific angular momentum of the particle, $L=b v^{(1)}_{\mathrm{K}}=b v^{(2)}_{\mathrm{K}}$ as the specific orbital angular momentum of the binary, and $a=a^{(1)} = a^{(2)}$.

Applying Newton's second law in spherical coordinates for the force terms derived from Eq.~\eqref{e:PNPhiInf} and  \eqref{e:PNXiInf}, we obtain:

\begin{equation}
  \ddot{r} = -\frac{M}{r^2} - \frac{3}{16} \frac{b^2 M}{r^4} + \frac{J^2}{r^3}
   + \frac{M J}{r^{4}}\left(2 a + \frac{L}{4}\right) + \mathcal{O}  \left(\frac{1}{r^5}\right)\, .
\end{equation}
The latter equation of motion can be derived from the effective potential:

\begin{equation}
  \Phi_{\mathrm{Eff}} =  -\frac{M}{r} - \frac{1}{16} \frac{b^2 M}{r^3} + \frac{J^2}{2r^2}
   + \frac{M J}{3r^{3}}\left(2 a + \frac{L}{4}\right) + \mathcal{O}  \left(\frac{1}{r^4}\right)\, .
  \label{e:PhiEff}
\end{equation}


\bibliography{sample63}{}
\bibliographystyle{aasjournal}



\end{document}